%% file: main.tex
\documentclass[11pt]{article}
\onecolumn % for one column layouts

\usepackage{amsmath,
amsfonts,
amsthm, 
mathabx,
bm,
mathrsfs,
graphicx,
enumerate,
fancyhdr,
mathtools, 
multicol, 
amssymb, 
caption, adjustbox, multirow, listings, graphicx}
\usepackage[table]{xcolor}
\definecolor{lightgray}{gray}{0.9}
\usepackage{enumerate}
\usepackage{natbib}
\usepackage{url} % not crucial - just used below for the URL 
\usepackage[margin=1in]{geometry}
\usepackage{algorithm}
\usepackage{algpseudocode}
\usepackage{booktabs,arydshln}
\usepackage{adjustbox}
\usepackage{ulem}
\usepackage{amsmath}
\usepackage{dsfont}
\usepackage{authblk}
\usepackage{graphicx}
\usepackage{booktabs, array, tabularx}
\usepackage[colorlinks=true,
            linkcolor=black,
            citecolor=blue,
            urlcolor=blue]{hyperref}

\theoremstyle{thmstyleone}%
\newtheorem{theorem}{Theorem}%  meant for continuous numbers
\newtheorem{proposition}[theorem]{Proposition}%
\theoremstyle{thmstyletwo}%
\theoremstyle{thmstylethree}%
\newtheorem{definition}{Definition}
\newtheorem{assumption}{Assumption}
\graphicspath{{Fig/}}

\begin{document}

\def\spacingset#1{\renewcommand{\baselinestretch}%
{#1}\small\normalsize} \spacingset{1}
\newcommand{\falco}[1]{{\color{red} #1}}
\newcommand{\vero}[1]{{\color{magenta} #1}}
\newcommand{\giulio}[1]{{\color{blue} #1}}
\newcommand{\leo}[1]{{\color{orange} #1}}

%\DOI{DOI HERE}
%\pubyear{2019}
%\access{Advance Access Publication Date: Day Month Year}
%\appnotes{Paper}

%\subtitle{Subject Section}

\title{Estimating the Effects of Heatwaves on Health: \linebreak A Causal Inference Framework}
\author[1]{Giulio Grossi\thanks{Corresponding author: giulio.grossi@unifi.it}}
\author[2]{Leo Vanciu}
\author[3]{Veronica Ballerini}
\author[3]{Danielle Braun}
\author[4,3]{Falco J.\ Bargagli-Stoffi}
\affil[1]{Department of Statistics, Computer Science, Applications, University of Florence}
\affil[2]{Faculty of Arts \& Sciences, Harvard University}
\affil[3]{Harvard T.H. Chan School of Public Health}
\affil[4]{Department of Biostatistics, University of California, Los Angeles}
\date{}

\maketitle

%\received{Date}{0}{Year}
%\revised{Date}{0}{Year}
%\accepted{Date}{0}{Year}

%\editor{Associate Editor: Name}

%\abstract{
%\textbf{Motivation:} .\\
%\textbf{Results:} .\\
%\textbf{Availability:} .\\
%\textbf{Contact:} \href{name@email.com}{name@email.com}\\
%\textbf{Supplementary information:} Supplementary data are available at \textit{Journal Name}
%online.}
 
% 200 words maximum for abstract: currently (11/13/2025)  192 words.
% Add one sentence on application when it is done.

\begin{abstract}
    The harmful relationship between heatwaves and health has been extensively documented in medical and epidemiological literature.
However, most evidence is associational and cannot be interpreted causally unless strong assumptions are made. 
In this paper, we first make explicit the assumptions underlying the statistical methods frequently used in the heatwave literature and demonstrate when these assumptions might break down in heatwave contexts.
To address these shortcomings, we propose a causal inference framework that transparently elicits causal identification assumptions.
Within this new framework, we first introduce synthetic controls (SC) for estimating heatwave effects, then propose a spatially augmented Bayesian synthetic control (SA-SC) method that accounts for spatial dependence and spillovers. 
Empirical Monte Carlo simulations show both methods perform well, with SA-SC reducing root mean squared error and improving posterior interval coverage under spillovers and spatial dependence.
Finally, we apply the proposed methods to estimate the causal effects of heatwaves on Medicare heat-related hospitalizations among 13,753,273 beneficiaries residing in Northeastern U.S. from 2000 to 2019.
This causal inference framework provides spatially coherent counterfactual outcomes and robust, interpretable, and transparent causal estimates while explicitly addressing the unexamined assumptions in existing methods that pervade the heatwave effect literature.
\end{abstract}

\vspace{0.5em}

\newpage
\section{Introduction}\label{sec:intro}

Recent global data indicate that billions of individuals experience prolonged periods of historically unprecedented high temperatures. 
Between 2023 and 2024, approximately 6.3 billion people—roughly 78\% of the global population—experienced at least 31 days with temperatures exceeding 90\% of those recorded in their local area during the 1991–2020 period \citep{GiguereEtAl2025ExtremeHeat}. 
Heatwaves, defined as prolonged periods of excessively high temperatures relative to local norms, are becoming more frequent, intense, and longer-lasting \citep{domeisen2023prediction}, posing significant public health risks worldwide \citep{faurie2022association}. %ye2012ambient

The health consequences of extreme heat manifest through multiple clinical pathways. 
Current estimates attribute hundreds of thousands of annual deaths globally to heat exposure, with elevated risk among elderly populations and individuals with comorbidities \citep{zhao2021global}.%\citep{analitis_effects_2014} 
Documented associations between heatwave exposure and adverse health effects include hyperthermia and heat exhaustion \citep{heo_temporal_2021}, respiratory complications \citep{lavigne_extreme_2014}, cardiovascular emergencies \citep{heo_temporal_2021}, emergency department surges \citep{nori-sarma_association_2022}, and hospital admissions across multiple diagnostic categories \citep{bobb_cause-specific_2014}.%bargagli2025extreme

Despite the extensive documentation of these associations, most existing evidence cannot be interpreted in a causal sense unless the plausibility of the underlying assumptions is supported. 
Many statistical approaches adopted in the literature to assess the relationship between heatwave exposure and health fail to (i) make such assumptions explicit, and (ii) verify their plausibility, impacting the interpretation of the results.
Prominent examples include distributed lag non-linear models (DLNM) \cite[see, e.g.,][]{gasparrini2011impact, guo2011tianjin}, generalized estimating equations (GEE) \cite[see, e.g.,][]{hajat2006impact, dippoliti2010impact}, mixed effect Poisson models \cite[see, e.g.,][]{anderson2009weather,toloo_sociodemographic_2014, guo2017heat}, hierarchical Bayesian models \cite[see, e.g.,][]{baccini2008heat,anderson2011heat}, matching procedures either on the heatwave exposure \cite[e.g.,][for temporal matching]{bobb_cause-specific_2014} or on the outcome \cite[e.g.,][for case-crossover design]{medina_extreme_2006}.%bargagli2025extreme 

The fundamental challenge for the causal interpretation lies in constructing valid counterfactuals that answer the question: ``\textit{what would have been the health of exposed populations (e.g., the number of hospitalizations, deaths, ...) in the absence of the heatwave?}''
To answer this question, one should refer to the concept of randomization: a scenario where, thanks to the absence of confounding (i.e., \textit{ignorability}), the only factor to which the effect can be attributed is the treatment (or \textit{exposure}) itself---in this case, the heatwave. 
When randomization is not feasible, as it is the case in most epidemiological and environmental studies, it should be emulated to draw causal conclusions.
Once \textit{quasi-randomization} is achieved under explicit assumptions, the target \textit{causal estimand}---the causal quantity of interest---can be identified. 
To our knowledge, this is seldom the case in the existing epidemiological literature on heatwaves.
However, recreating randomization becomes particularly complex in the heatwave framework, given the spatio-temporal nature of the exposure.

%This paper makes three contributions to the literature. 
Our contribution is threefold.
First, we provide a comprehensive critical analysis of existing statistical methods---including DLNMs, Bayesian hierarchical methods, and matching designs---elucidating their implicit causal assumptions, clarifying what their target estimands are, and demonstrating where these assumptions might break down under realistic applied scenarios (Section \ref{sec:background}). 
Particular attention is given to the spatial and temporal structure of the data. 
In fact, from the spatial perspective, geographic proximity induces shared vulnerability and potential spillover effects, where neighboring areas experience similar weather patterns and may see population movement during extreme heat events \citep{wellenius_heat-related_2017}. 
This spatial structure, if not explicitly accounted for, can bias the model estimates \citep{papadogeorgou2019adjusting}. 
From a temporal perspective, adopted methods, such as temporal matching, where the same unit is compared across different time periods, do not account for time-varying confounders and neglect potential spillovers between neighboring units.
Addressing these methodological challenges requires rigorous causal inference frameworks that explicitly incorporate spatial structure while accounting for the time-varying component of heatwave exposures. 
Existing \textit{quasi-experimental} approaches typically assume independence between units---the so-called \textit{no-interference} assumption \citep{forastiere2021identification,bargagli2025heterogeneous}---an assumption often violated in spatially correlated heatwave exposures. 

Second, building on these identified limitations, we introduce a causal inference framework that transparently and explicitly targets the identification and estimation of heatwave effects (Section \ref{sec:method}). 
We start by introducing a framework that explicitly accounts for the spatial component of heatwaves in the identification of their causal effects. Within this framework, we first introduce synthetic controls (SC) for estimating heatwave effects, then propose a spatially augmented Bayesian synthetic control (SA-SC) method that accounts for spatial dependence and spillovers. 
The proposed methods are flexible semiparametric approaches---building on the original SC approach \citep{abadie2010synthetic, martinez2022bayesian}---that integrate geographic distance into SC estimation to construct reliable control units, such as exposed populations in the absence of the heatwave. 
The proposed methods innovate the methodological and applied literature by: 
(i) transparently eliciting the assumptions needed to causally identify the effect of heatwaves on health (SC); 
(ii) explicitly incorporating the spatial structure to account for spatial dependence and spillover (SA-SC). 
We argue that these advances provide a more realistic representation of the underlying process in heatwave studies. 

Third, we evaluate the proposed methods through two complementary data applications. 
First, we conduct extensive empirical Monte Carlo simulations using real-world meteorological data from the GridMET dataset, which provides daily estimates of temperature, precipitation, and humidity at $4 \times 4$ km resolution across the Northeastern United States from 2000 to 2019 \citep{abatzoglou2013development}. We simulate health outcomes mimicking real-world distributions and test both SC and SA-SC under violations of the no interference and no spatial dependence assumptions (Section \ref{sec:simulation}). 
Second, we apply these methods to actual heat-related hospitalizations in the same region and period. 
We obtained hospitalization data from the Centers for Medicare \& Medicaid Services (CMS), which maintains comprehensive administrative claims records for all Medicare beneficiaries in the United States. 
Medicare is the national health insurance program covering Americans aged 65 and older, representing a population particularly vulnerable to heat-related health impacts.
We aggregate hospitalizations to the county-day level and classify them by principal diagnosis using International Classification of Diseases (ICD) codes which identify diagnoses recorded at discharge, harmonized across revisions using standard crosswalks. 
Diagnostic groups include heat-related illness, fluid and electrolyte disorders, acute kidney failure, urinary tract infection, and septicemia. 
For each urban county-day, we compute hospitalization rates per 100,000 Medicare fee-for-service beneficiaries (Section \ref{sec:application}).
To ensure dissemination and facilitate public access, we provide code for each step of the analysis.

Finally, a discussion of the broader implications for heatwave policy, methodological trade-offs, and future research directions is provided in Section \ref{sec:discussion}. Appendix Table \ref{tab:notation} summarizes the full notation used throughout the paper to assist readers in following the methodological development and empirical analyses.

\vspace{-0.5cm}

\section{Background}\label{sec:background}

\subsection{\textbf{Setup}}

Let $Y_{it}$ denote a health outcome for (areal) unit $i$ (with $i=1,\dots,N$) at time $t$ (with $t=1,\dots,T$).
In this work's case study, units may represent cities, ZIP codes, counties, or other administrative regions; data frequency is typically daily. 
Let $\mathbf{X}_{it}$ be a $p-$dimensional vector of covariates associated with each unit (e.g., demographics, socio-economic status, and other characteristics). 
Throughout, uppercase letters will be used to denote random variables, while lowercase letters will represent realized values of those variables. Bold letters represent vectors and matrices.

The epidemiological literature employs various definitions of heatwaves, typically characterized by temperature thresholds and duration requirements \citep{anderson2011heat, bobb_cause-specific_2014, gasparrini2011impact}.
\begin{definition}[Heatwave]
\textit{Let $H_{it}$ denote the daily mean temperature for unit $i$ at time $t$.}
\footnote{\scriptsize Some studies employ absolute temperature thresholds \cite[e.g.,][]{baccini2008heat} or heat index which combines temperature and humidity \citep{anderson2011heat}.} 
\textit{Let $q_i^r$ be the $r$-th percentile of the historical temperature distribution for unit $i$ during a certain time period.}
\footnote{\scriptsize Usual reference periods for the construction of percentiles are warm season, year, or study period \citep{bobb_cause-specific_2014}.} 
\textit{A heatwave occurs in unit $i$ starting at time $T_0$ if $H_{it} \geq q_i^r$ for all $t \in [T_0, T_0 + \ell]$ (intensity criterion), and where $\ell$ is at least equal to a pre-specified number of consecutive days (duration criterion).
The heatwave treatment indicator is defined as: $Z_{it} = 1$ if unit $i$ is experiencing a heatwave at time $t$ and $0$ otherwise.
The indicator $Z_{it}$ is time-varying and can switch between 0 and 1 multiple times within a season, reflecting the transient nature of heatwave events.}
\footnote{\scriptsize Several studies exclude the first day of a heatwave when estimating the heatwave effects, on the premise that physiological stress accumulates over multiple days of exposure \citep{bobb_cause-specific_2014}.}
\end{definition}
 Common operational definitions in the literature use percentile thresholds ranging from the 90th to 99th percentiles of location-specific temperature distributions \citep{gasparrini2011impact, anderson2009weather} with duration requirements of 2 or more consecutive days \citep{guo2017heat}. 
For our application, we adopt the widely-used definition of temperatures exceeding the 95th percentile for at least two consecutive days
\citep{bobb_cause-specific_2014}. 
However, our method can be applied to any definition of heatwave.

Under the potential outcomes framework and the Stable Unit Value Assumption \citep[SUTVA;][]{rubin1978}---which implies no spillovers, i.e., $Y_{it}(\mathbf{Z}) = Y_{it}(Z_{i})$, and \textit{consistency}, i.e., $Y_{it} = Y_{it}(z)$ if $Z_{it} = z$---we postulate the existence of two potential outcomes: $Y_{it}(1)$, which denotes the health outcome for unit $i$ at time $t$ under heatwave exposure, and $Y_{it}(0)$, which denotes the health outcome for unit $i$ at time $t$ under no heatwave. 

The causal effect of heatwave exposure can be quantified through different comparisons of $Y_{it}(1)$ and $Y_{it}(0)$, defining different causal estimands depending on the outcome type and research question.

For count outcomes (e.g., mortality counts, hospital admissions), the causal estimand of interest---which is the target parameter---may be expressed on the additive scale as $\mathbb{E}[Y_{it}(1) - Y_{it}(0)]$, representing excess events, or on the multiplicative scale as $\mathbb{E}[Y_{it}(1)]/\mathbb{E}[Y_{it}(0)]$, representing relative risk. 
When $Y_{it}$ represents rates (e.g., deaths per 100,000 population), these estimands capture excess mortality rates and rate ratios, respectively, providing population-adjusted measures that are comparable across geographic units of different sizes. 

The multiplicative scale is typically preferred for count and rate data as it accounts for baseline variations across units, maintains non-negativity, and provides interpretable percent changes. 
Furthermore, the majority of studies in the heatwave effects literature report results in the multiplicative scale \cite[see, e.g.,][]{hajat2006impact, gasparrini2011impact, guo2011tianjin, bobb_cause-specific_2014, medina-ramon_temperature_2007}.% bargagli2025extreme 
Thus, throughout the paper, we will focus on multiplicative scale estimands.

\subsection{\textbf{State-of-the-art Methods in the Heatwave Effects Literature}}

Statistical methods in heatwave epidemiology have provided valuable insights into heatwave-health relationships through various modeling approaches. 
In particular, generalized estimating equations (GEE) \cite[see, e.g.,][]{hajat2006impact, dippoliti2010impact, anderson2009weather}, overdispersed Poisson and negative binomial models \cite[see, e.g.,][]{anderson2009weather,toloo_sociodemographic_2014, guo2017heat}, distributed lag non-linear models (DLNM) \cite[see, e.g.,][]{gasparrini2010distributed,gasparrini2011impact, guo2011tianjin}, hierarchical Bayesian models \cite[see, e.g.,][]{baccini2008heat,anderson2011heat}, and matching procedures either on the heatwave exposure \cite[e.g.,][for temporal matching]{bobb_cause-specific_2014} or on the outcome \cite[e.g.,][for case-crossover design]{medina_extreme_2006} have been adopted in the literature.%bargagli2025extreme

We systematically review these seminal methods, explicating their statistical parameters and assumptions, and discussing what would be required for formal causal interpretation. We would like to stress that, while we explicitly elicit the assumptions needed for causal identification (and formally introduce causal identification of such models in the Appendix), such assumptions are often unexamined in the heatwave effect literature.

\subsubsection{Regression-Based Multi-Community Studies}\label{ss:gee}

Multi-community studies \citep{hajat2006impact,anderson2009weather,dippoliti2010impact} employ generalized estimating equations (GEE) or mixed-effects Poisson regression to estimate community adjusted associations: %analitis_effects_2014
$$\begin{aligned}
    \log(\mathbb{E}[Y_{it}]) = \bm{\alpha}_{\text{C}} + \beta Z_{it} + \mathbf{X}_{it}'\boldsymbol{\gamma}
\end{aligned}$$
where $\bm{\alpha}_{\text{C}}$ represents community-level fixed effects---$\bm{\alpha}_C$ takes the same value for all $i$ in the same community, i.e., $C_i = c$. 
Communities can be regions, provinces, cities, and so on.
GEE uses a working correlation matrix to account for within-community temporal dependence. 
The parameter $\exp(\beta)$ represents the multiplicative association between heatwaves and health outcomes, pooled across communities while adjusting for community-specific baseline rates.

For causal interpretation of $\exp(\beta)$ as a community-adjusted relative risk $\text{RR}^{\text{GEE}} = \mathbb{E}[Y_{it}(1) \mid C_i]/\mathbb{E}[Y_{it}(0) \mid C_i]$, one would require: 
(i) SUTVA both within and across communities; 
(ii) conditional ignorability: $(Y_{it}(1), Y_{it}(0)) \perp\!\!\!\perp Z_{it} \mid C_i, \mathbf{X}_{it}$; and (iii) treatment overlap within each community: $0 < P(Z_{it}=1 \mid C_i, \mathbf{X}_{it}) < 1$. 
Additionally, for pooling across communities, one needs either effect homogeneity or a correctly specified model for effect heterogeneity.
When overdispersion is present, quasi-Poisson or negative binomial specifications may be employed \citep{anderson2009weather,toloo_sociodemographic_2014, guo2017heat}, though the interpretation remains unchanged. 
Identification results under the listed identifying assumptions are provided in the Supplementary Materials.

\subsubsection{Hierarchical Bayesian Models}\label{ss:bayes}

Hierarchical Bayesian approaches model community-specific associations \citep{dominici2000combining}, explicitly assuming $Y_{it} \sim \text{Poisson}(\zeta_{it})$, and
$$\begin{aligned}
     \log(\zeta_{it})  = \alpha_i + \beta_i Z_{it} + v(\mathbf{X}_{it}; \boldsymbol{\gamma}_i), \\ 
     \beta_i \sim \text{N}(\mu, \sigma^2)
\end{aligned}
$$
here $v(\cdot)$ represents smooth functions of confounders including temperature and temporal trends \citep{bobb2011bayesian,anderson2011heat}. 
\cite{bobb2011bayesian} employ Bayesian model averaging over different specifications of $v(\cdot)$ to account for model uncertainty. 
\cite{anderson2011heat} extend this framework by allowing $\beta_i$ to depend on heatwave characteristics through $\beta_i \sim \text{N}(\alpha_0 + \alpha_1 \mathbf{W}_i, \sigma^2)$, where $\mathbf{W}_i$ includes intensity, duration, and timing.

The posterior distribution of $\exp(\beta_i)$ provides community-specific association measures. For causal interpretation as relative risks $\text{RR}_i^{\text{HB}} = \mathbb{E}[Y_{it}(1) \mid \alpha_i]/\mathbb{E}[Y_{it}(0) \mid \alpha_i]$, one needs: 
(i) SUTVA at the community level; 
(ii) conditional ignorability: $(Y_{it}(1), Y_{it}(0)) \perp\!\!\!\perp Z_{it} \mid \alpha_i, \mathbf{X}_{it}$; 
(iii) treatment overlap: $0 < P(Z_{it}=1 \mid \alpha_i, \mathbf{X}_{it}= \mathbf{x}) < 1$.
The random effects $\alpha_i$ and covariates must capture all confounders. Identification results under these assumptions are provided in the Supplementary Materials.

\subsubsection{Distributed Lag Non-Linear Models (DLNM)}\label{ss:dlnm}

Distributed Lag Non-linear Models (DLNM) methods have been widely adopted in the epidemiological literature \cite[see, e.g.,][and subsequent works]{gasparrini2010distributed,guo2011tianjin}.
%gasparrini2010distributed, guo2011tianjin, gasparrini2011impact,gasparrini2014modeling,gasparrini2015mortality
DLNMs estimate associations between heatwaves and health outcomes using quasi-Poisson regression with distributed lag structures:
$$    \log(\mathbb{E}[Y_{it}]) = \alpha + h(H_{it}, \text{lag}) + \beta Z_{it} + g(t) + \mathbf{X}_{it}'\boldsymbol{\gamma}
$$ 
where $h(\cdot)$ is a cross-basis function capturing non-linear and delayed temperature effects, $g(\cdot)$ is a spline function, and $\mathbf{X}_{it}$ also includes meteorological and seasonal terms. 
The parameter $\exp(\beta)$ quantifies the multiplicative association between heatwave exposure and health outcomes, often interpreted as an ``added effect'' beyond individual hot days. 

For one to interpret the $\beta$ parameter from a causal perspective---targeting an estimand such as the conditional relative risk (RR), $\text{RR}^{\text{DLNM}} = \mathbb{E}[Y_{it}(1) \mid \mathbf{X}_{it}]/\mathbb{E}[Y_{it}(0) \mid \mathbf{X}_{it}]$---one would need to invoke the following assumptions: 
(i) SUTVA; 
(ii) conditional ignorability: $(Y_{it}(1), Y_{it}(0)) \perp\!\!\!\perp Z_{it} \mid H_{it}, g(\text{t}), \mathbf{X}_{it}$; and 
%requiring all confounders to be measured; and 
(iii) treatment overlap: $0 < P(Z_{it}=1 \mid H_{it}, g(\text{t}), \mathbf{X}_{it}) < 1$. %, requiring that each geographical unit has a positive probability of being exposed to a heatwave. 
Under these conditions, $\exp(\beta)$ would identify the causal relative risk.
Identification results under the listed identifying assumptions are provided in the Supplementary Materials.

\subsubsection{Matching Designs}\label{ss:cc}

Matching designs in heatwave epidemiology compare the same units across different time periods.
This includes comparing the same units on days with specific exposure characteristics to days without those characteristics, such as comparing heatwave to non-heatwave days to estimate temperature-related mortality \citep{bobb_cause-specific_2014}. 
Alternatively, \textit{case-crossover designs} match outcome events within the same individual across different time periods, comparing exposures when an outcome occurred to exposures during control periods \citep{medina-ramon_temperature_2007,guo2011tianjin}. 
In the following, we will focus on \textit{exposure-based temporal matching methods} like those comparing heatwave to non-heatwave days, though the matching principles can be applied to case-crossover designs too.

%Matching designs compare either exposures during the event period to exposures during control periods within the same unit, effectively using each unit as its own control \citep{bobb_cause-specific_2014}, \falco{or match outcome events with non-outcome events to assess the influence of the exposure on the outcome \citep{medina-ramon_temperature_2007,guo2011tianjin}.
%In the following, we will discuss temporal matching methods., but the same arguments can be easily translated to the case of outcome/non-outcome matching (which is also referred to as \textit{case-crossover design} in the literature).}
%These are implemented either at the area level comparing the same geographic unit across time periods, or at the individual level comparing the same person's exposure at different times. 

In the context of heatwaves, formally, for unit $i$, exposure-based temporal matching methods compare $Y_{it}$ during heatwave days ($Z_{it}=1$) with $Y_{it'}$ during matched control days ($Z_{it'}=0$) from previous weeks or years, typically using conditional Poisson regression:
$$ \log(\mathbb{E}[Y_{it}]) = \alpha_i + \beta Z_{it}$$
where matched periods share calendar characteristics (day of week, season) and $\alpha_i$ represents unit-specific intercepts. 

To interpret $\exp(\beta)$ causally as a unit-specific relative risk $\text{RR}^{\text{CC}}_i = \mathbb{E}[Y_{it}(1)]/\mathbb{E}[Y_{it'}(0)]$, one requires: 
(i) SUTVA; 
(ii) within-unit temporal exchangeability: $\mathbb{E}[Y_{it}(0) \mid X_i=x] = \mathbb{E}[Y_{it'}(0) \mid X_i=x]$ for matched control periods $t'$, implying no secular trends or time-varying confounders between matched time points;\footnote{\scriptsize To mitigate this issue, researchers have proposed a bi-directional matching procedure which uses, as control matched units, both past and future units \citep{bobb_cause-specific_2014}.} 
(iii) no carry-over effects: $Y_{it}(z_{it}) = Y_{it}(\mathbf{z}_{i})$ regardless of treatments at other times; 
(iv) treatment overlap;
and (v) stable unit composition across periods.
Identification results under the listed identifying assumptions are provided in the Supplementary Materials.

Some implementations model hospitalization or mortality rates rather than counts \citep{karlsson2018population}, using weighted least squares with population denominators. 
Others employ quasi-Poisson specifications to handle overdispersion \citep{schifano_susceptibility_2009}. 
While these variations may improve statistical efficiency or change the scale of the estimand (from counts to rates), they maintain the same fundamental causal structure.\footnote{\scriptsize Similar to case-crossover designs, \textit{case-only} approaches have also been adopted in the literature. 
These studies analyze only cases (e.g., deaths or hospitalizations) and test whether individual characteristics (age, sex, comorbidities) or contextual factors (poverty, air conditioning access) are differentially distributed among cases occurring on hot versus non-hot days \citep{madrigano_case-only_2015,schwartz_who_2005}.
Given the fact that such models do not directly assess the effects of heatwave on health outcomes, we have excluded them from the methods review.}

\subsection{\textbf{Challenges to Causal Inference in Heatwave Epidemiology}}

While these statistical approaches have generated critical associational empirical findings, formal causal interpretation requires, as we have described above, the introduction of identifying assumptions. 
Such assumptions, which are untestable, may generally be plausible after a careful study design phase \citep{imbens2015causal}.
\footnote{These assumption are untestable because they either refer to the potential oucomes, which are never jointly oberved, as in the case of conditional ignorability, or to the unknown treatment assignment mechanism, as in the case of treatment overlap.}
However, that may still be violated in epidemiological contexts, such that of heatwave effects assessment. 

For instance, conditional ignorability $(Y_{it}(1), Y_{it}(0)) \perp\!\!\!\perp Z_{it} \mid \mathbf{X}_{it}$ becomes questionable even after reaching a good balance of the observed covariates, due to the likely presence of unmeasured spatiotemporal confounders, such as air pollution co-occurring with heatwaves, population vulnerability shifts, or adaptive behaviors. 
Such \textit{unobservables} affect both heatwave occurrence and health outcomes \citep{papadogeorgou2019adjusting}, preventing the causal interpretation of the effects. 
In the case of time-varying spatial unmeasured confounders, not even fixed effects can control for them  \citep{gunasekara2014fixed}.

Yet, within-units matching methods (exposure-based temporal matching methods, case-cross-over designs), which were adopted to relax the reliance on conditional ignorability, effectively eliminate time-invariant confounding. 
However, these methods require temporal exchangeability---that is, analyzed units are not changing in time. Such assumption fails under secular mortality trends or climate adaptation \citep{benmarhnia2015vulnerability}.%achebak2019trends

Finally, and perhaps most critically, all approaches assume SUTVA, yet one can imagine heatwaves exposures violate this by various channels. 
Spatial spillovers can, in fact, manifest through: 
(i) population movement from affected to unaffected areas \citep{hondula_heat-related_2014}; 
(ii) shared infrastructure (e.g., local power grids) affecting multiple units \citep{stone2021compound}; 
(iii) emergency response reallocation across boundaries \citep{knowlton20092006}; and (iv) behavioral spillovers \citep{bargagli2025heterogeneous}. 
In the next section, we propose a novel causal inference framework to identify and estimate heatwave effects that tackles these complexities.

%\vspace{-0.25cm}
\section{Causal Inference Framework for Heatwave Effects Identification and Estimation}\label{sec:method}

Modern causal inference provides a rigorous framework that directly addresses two critical features of heatwave exposure: 
(i) it establishes transparent identification assumptions for constructing valid counterfactuals from observational data, enabling researchers to explicitly state and evaluate the conditions required for causal claims, 
and (ii) it accommodates the inherent spatial structure of heatwave events, offering principled methods to accommodate both spatial dependence and spillover effects. 

While these frameworks offer substantial advantages, their application to heatwave epidemiology requires addressing the fact that standard causal inference methods for temporal data face significant challenges when applied to heatwave effects. 
Difference-in-differences (DiD) approaches require parallel trends assumptions between treated and control units, which are often violated when geographic areas have heterogeneous baseline temperature-health relationships and differential adaptation capacities \citep{papadogeorgou2022causal,grossi2024direct,grossi2025spatial}. 
Interrupted time series designs assume a sharp, permanent treatment onset, yet heatwaves are transient events that can appear multiple times with potential lagged effects \citep{gasparrini2011impact,gasparrini2014modeling}. 
Time series fixed-effects models can account for time-invariant confounders but struggle with time-varying factors like changing population demographics, evolving heat adaptation measures, and concurrent environmental exposures \citep{deschenes2011climate}. 
Instrumental variable approaches are limited by the absence of plausible instruments that affect heatwave exposure but not health outcomes directly \citep{hsiang2016climate}. 
Moreover, many of these methods typically assume homogeneous treatment effects across units which is often unrealistic given varying local vulnerabilities \citep{heutel2021adaptation}.

Against this backdrop, the synthetic control (SC) method \citep{abadie2010synthetic} offers a principled alternative that addresses these limitations by constructing unit-specific counterfactuals while being transparent about the required identifying assumptions. 
In Section \ref{ss:synth}, we adapt the SC approach to the setting of heatwave effects in temporal data. 
While SC has been widely used in policy evaluation, its application to heatwave exposures with multiple treated units represents a novel extension.\footnote{\scriptsize Although SC has been used in other environmental applications \cite[see, e.g.,][]{sills2015estimating,andersson2019carbon,huang2022use}.} 
Furthermore, we explicitly articulate the assumptions required for causal identification in this setting.
Second, in Section \ref{ss:sasynth}, we extend the SC framework to fully spatio-temporal data by developing spatially-augmented Bayesian SC methods. 
This method explicitly incorporates geographic dependencies that standard SC methods ignore and, in turn, accounts for the reality that neighboring areas experience similar weather patterns and potential spillover effects.
Our spatial augmentation addresses these violations while maintaining the SC framework's transparency regarding identification assumptions, providing a more complete framework for spatio-temporal heatwave data.

\subsection{\textbf{Causal Inference with Temporal Data: Applying Synthetic Control Methods to Heatwave Effects}}\label{ss:synth}

The seminal SC method \citep{abadie2010synthetic} estimates the counterfactual outcome for a treated unit as a convex combination of untreated ``donor'' units chosen to reproduce the pre-treatment behavior of the outcome. 
Let $\mathcal{Z}_t = \{i : Z_{it} = 1\}$ denote the set of all units experiencing a heatwave at time $t$, and $\mathcal{N}_0^t = \{i : Z_{it} = 0\}$ denote the potential donor pool at time $t$.
For each treated unit $i \in \mathcal{Z}_t$, we construct a separate synthetic control using the donor pool $\mathcal{N}_0^t$.

Consider a specific treated unit $i \in \mathcal{Z}_t$ that experiences a heatwave after period $T_0$.
A standard structural justification adopts an interactive fixed-effects representation for potential outcomes under no heatwave exposure:
\begin{equation}\label{eq:dgp}
    Y_{it}(0)=\mu_i+\delta_t+f_t^\top \lambda_i+\varepsilon_{it},
\end{equation}
where $\delta_t$ captures common temporal shocks (e.g., seasonal mortality patterns), $\mu_i$ unit-specific effects (e.g., baseline health infrastructure), $f_t\in\mathbb{R}^r$ time-varying common factors, $\lambda_i\in\mathbb{R}^r$ unit-specific loadings \footnote{\scriptsize A factor loading represents the strength and direction of the relationship between an observed variable and an underlying latent factor, indicating how much the variable is influenced by that common component.} and $\varepsilon_{it}$ idiosyncratic errors with mean zero.\footnote{\scriptsize For compact notation, we do not index this and the following error terms relative to the controls with $(0)$---e.g., $\varepsilon_{it}^{(0)}$.}
For treated unit $i$, let $w=(w_j)_{j\in\mathcal{N}_0^t}$ be a vector of weights that lie in the simplex:
$$\Delta^{|\mathcal{N}_0^t|-1}=\Big\{\,w\in\mathbb{R}^{|\mathcal{N}_0^t|}:\ w_j\ge 0,\ \sum_{j\in\mathcal{N}_0^t} w_j=1\,\Big\}.
$$
SC chooses weights by minimizing the pre-heatwave discrepancy between the treated unit's outcomes and the weighted combination of control outcomes,
\begin{equation}\label{eq:weights}
\widehat{w}^{i} \in \arg\min_{w\in\Delta^{|\mathcal{N}_0^t|-1}}
\sum_{t\le T_0}\Big(Y_{it}-\sum_{j\in\mathcal{N}_0^t} w_j Y_{jt}\Big)^2.
\end{equation}
The estimated counterfactual for post-heatwave periods is then
$$\widehat{Y}_{it}(0)=\sum_{j\in\mathcal{N}_0^t}\widehat{w}_j^i\,Y_{jt}\;, \qquad t>T_0.$$
The estimation of a credible synthetic control for heatwave effects relies on a set of assumptions that should be considered carefully in the heatwave context.

\begin{assumption}[\textbf{Weights Interpolation}  \citep{abadie2010synthetic}]\label{ass:overlap}
\textit{There exist weights
$w\in\Delta^{|\mathcal{N}_0^t|-1}$ such that
$\lambda_i=\sum_{j\in\mathcal{N}_0^t} w_j \lambda_j$.}
\end{assumption}
\noindent Weights interpolation requires that the heatwave-exposed unit's factor loadings are (approximately) interpolable by those of the non-exposed donors. Under weights interpolation, the counterfactual discrepancy decomposes as
$$
Y_{it}(0)-\sum_{j\in\mathcal{N}_0^t} w_j Y_{jt}(0)
=\Big(\mu_i-\!\sum_j w_j\mu_j\Big)
+f_t^\top\!\Big(\lambda_i-\!\sum_j w_j \lambda_j\Big)
+\varepsilon_{it}-\!\sum_j w_j\varepsilon_{jt},
$$
thus reducing pre-heatwave imbalance in $\bm \mu$ and $ \bm \lambda$ controls the post-heatwave systematic error provided the factor structure is stable.
Weights interpolation is more plausible when the heatwave-exposed area's climate profile, demographic composition, baseline mortality rates, and health-system capacity can be credibly approximated by a convex combination of non-exposed areas---for instance, when donor units---units not exposed to treatment neither in the pre-treatment nor in the post-treatment---share similar characteristics and baseline heat vulnerability. 
The assumption becomes fragile with structural heterogeneity such as recent heat adaptation programs, changes in emergency response protocols, or local health infrastructure improvements that push $\lambda_i$ outside the donors' convex hull.

\begin{assumption}[\textbf{Stable Pattern} \citep{abadie2010synthetic, xu2017generalized}]\label{ass:stablepattern}
\textit{
Let potential outcomes under no heatwave exposure satisfy equation \ref{eq:dgp} with time-varying common factors $\{f_t\}$ and unit loadings $\{\lambda_i\}$.
We assume that the same interactive factor structure that governs the pre-heatwave period continues to hold post-heatwave for the untreated potential outcomes, and that no unit-specific structural break affects the treated unit alone:
$$
\begin{aligned}
&\big\{(\lambda_i,f_t,\mu_i,\delta_t): t\le T_0\big\}
\ \stackrel{d}{=}\ 
\big\{(\lambda_i,f_t,\mu_i,\delta_t): t> T_0\big\}
\quad\text{for } i\in\mathcal{Z}_t \cup\mathcal{N}_0^t,\\
&(\lambda_i,f_t,\mu_i)\ \text{is stationary for } i \in \mathcal{Z}_t.    
\end{aligned}$$
}
\end{assumption}

This assumption is plausible when the determinants of mortality and morbidity that are common across geographic units (e.g., seasonal disease patterns, long-run demographic trends, baseline temperature-mortality relationships) evolve smoothly through $T_0$, and when no location-specific regime change occurs exactly at heatwave onset. 
This assumption would be violated if, for example, the heatwave-exposed area simultaneously implemented new heat emergency protocols not adopted elsewhere, or if hospital capacity constraints uniquely affected the treated location.

\begin{assumption}[\textbf{No Anticipation} \citep{abadie2021}]\label{ass:noanticipation}
\textit{We assume that the heatwave has no effect before $T_0$: $Y_{it}(1)=Y_{it}(0) \:\: \forall  \:\: t \le T_0$ and $i \in \mathcal{Z}_t$.}
\end{assumption}
If the no anticipation assumption fails and there exist nonzero anticipatory effects $\tau_{it}^{\mathrm{lead}}$ on a subset $\mathcal{T}_{\mathrm{lead}}\subseteq\{1,\dots,T_0\}$, then the SC objective is minimized on a shifted target, leading to biased weight estimation.
Assumption \ref{ass:noanticipation} is credible when behavioral and institutional responses that could affect health outcomes (e.g., heat warnings, cooling center openings, hospital surge protocols, public health messaging) do not activate before the formal heatwave onset at $T_0$. 
The assumption becomes more fragile when meteorological forecasts trigger early interventions---for instance, if heat advisories issued days before the actual heatwave lead to preemptive hospitalizations or changes in healthcare-seeking behavior, shifting $Y_{it}$ in the late pre-period and biasing the synthetic control weights.\footnote{\scriptsize This assumption is supported by evidence that people do not substantially adjust their behavior even when heat warnings are issued \citep{toloo2013evaluating}. 
As further evidence, \cite{weinberger2021heat} found no significant mortality reductions from heat alerts in most U.S. cities studied.}
\begin{assumption}[\textbf{Idiosyncratic Shocks} \citep{abadie2010synthetic}]\label{ass:shocks}
\textit{For all $t$, $\mathbb{E}\!\left[\varepsilon_{it}-\sum_{j\in\mathcal{N}_0^t} -\hat{w_j}\,\varepsilon_{jt}\right]=0.$}
\end{assumption}
Under SUTVA and assumptions \ref{ass:overlap}--\ref{ass:shocks}, SC can be used to identify the causal effect of heatwaves. Our causal estimand of interest is the Relative Risk for the Treated (RRT) during the post-treatment period:
\begin{equation}\label{eq:rr-att}
    \text{RRT}^{\text{SC}} = \frac{\mathbb{E}[Y_{it}(1) \mid i \in \mathcal{Z}_t,  T_0 < t < T_0+\ell]}{\mathbb{E}[Y_{it}(0) \mid i \in \mathcal{Z}_t, T_0 < t < T_0+\ell]}
\end{equation}
which quantifies the average multiplicative causal effect of heatwaves on health outcomes across all treated units the during heatwave.\footnote{\scriptsize One could also examine lagged effects by extending the post-treatment window 
beyond the heatwave duration, i.e., $T_0 < t < T_0 + \ell + \psi$ where $\psi$ represents the lag period.} 
The numerator represents observed outcomes for heatwave-exposed units, while the denominator represents the counterfactual outcomes that would have been observed in the absence of heatwaves, estimated via the SC as $\widehat{Y}_{it}(0) = \sum_{j\in\mathcal{N}_0^t} \widehat{w}_j^i Y_{jt}$ for each treated unit $i \in \mathcal{Z}_t$. Identification results under the listed identifying assumptions are provided in the Supplementary Materials.

\subsection{\textbf{Causal Inference with Spatio-Temporal Data: Identification Under Partial Interference}}\label{ss:sasynth}

The SC approach outlined in Section \ref{ss:synth} assumes SUTVA, which requires that a unit's potential outcome depends only on its own treatment status. 
However, as argued above, this assumption is likely violated in heatwave contexts due to spatial spillovers between adjacent geographic units.
In fact, heatwaves are meteorological phenomena that typically affect spatially contiguous regions. This empirical regularity that extreme heat events span connected geographic areas rather than isolated points motivates our modeling approach for spatial interference. 
To address this challenge, inspired by \cite{forastiere2021identification}, we relax SUTVA and adopt a modified Stable Unit Treatment Value Assumption on Neighbors (SUTNVA), which implies a different potential outcomes' definition.

\begin{assumption}\label{ass:sutnva}[\textbf{SUTNVA -- Stable Unit Treatment Value Assumption on Neighbors}]\label{ass:sutnva}
\textit{For unit $i$ at time $t$, the potential outcome can be written as $Y_{it}(Z_i, m(\mathbf{z}_{N_i^{(s_0)},t}))$ where $m(\cdot)$ is an exposure mapping function summarizing the treatment status of neighboring units. 
Here, $N_i^{(s_0)} = \{j : 1 \leq s_{ij} \leq s_0\}$ denotes the set of units within degree $s_0$ of unit $i$, where $s_{ij}$ represents the degree of separation ($s_{ij} = 1$ for contiguous units sharing a border, $s_{ij} = 2$ for units separated by one intermediate unit, etc.).
We adopt the exposure mapping $m(\mathbf{z}_{N_i^{(s_0)},t}) = \mathds{1}\{\sum_{j \in N_i^{(s_0)}} z_{jt} = |N_i^{(s_0)}|\}$, which equals 1 if all $s_0$-degree neighbors are treated and 0 otherwise.}
\end{assumption}

\begin{definition}[\textbf{Relevant Potential Outcomes Under Spatial Interference}]
\textit{Under SUTNVA with spatial heatwave exposure, we consider four potential outcomes. 
First, $Y_{it}(1,1)$ represents the outcome when unit $i$ and all its $s_0$-degree neighbors experience a heatwave. 
Second, $Y_{it}(0,0)$ represents the outcome when unit $i$ and none of its $s_0$-degree neighbors experience a heatwave. 
Third, $Y_{it}(0,1)$ represents the outcome when unit $i$ does not experience a heatwave but all its $s_0$-degree neighbors do. 
Fourth, $Y_{it}(1,0)$ represents the outcome when unit $i$ experiences heatwave but no neighbors do.}
\end{definition}

We note that, since heatwaves typically affect contiguous regions, we expect $Y_{it}(1,1)$ and $Y_{it}(0,0)$ to be the predominant configurations in the data.\footnote{\scriptsize This is, of course, dependent on the choice of $s_0$.} 
The configuration $Y_{it}(1,0)$ would require an isolated heatwave affecting a single unit, which is extremely rare for the spatial scale of our analysis (e.g., county-level). 
The configuration $Y_{it}(0,1)$ could occur at boundaries of heatwave regions but represents a small fraction of observations.
To ensure clean identification of causal effects under SUTNVA, we construct donor pools that account for potential spatial contamination:
\begin{definition}[\textbf{Spatially-Adjusted Donor Pool}]
\textit{For each treated unit $i \in \mathcal{Z}_t$ experiencing a heatwave starting at $T_0$, we define the donor pool as:
$$\mathcal{N}_0^{*t} = \{j : Z_{jt} = 0 \text{ and } \min_{i \in \mathcal{Z}_t} s_{ij} > s_0 \}.
$$
This construction ensures donor units are sufficiently distant to avoid both direct treatment and spillover effects.}\footnote{\scriptsize The choice of $s_0$ degrees separation provides a buffer zone for the treatment radius and to account for potential spillovers.}
\end{definition}
\noindent
Following \cite{papadogeorgou2019adjusting}, among eligible donors, we prioritize units at distance $s_0+1$ from treated units, as these are most similar climatologically while remaining uncontaminated from spillovers.

The synthetic control weights $w^i$ are then chosen as in Equation \eqref{eq:weights} for each treated unit $i$.
Under SUTNVA and assumptions 1-4, we define the the causal estimand of interest.
\begin{definition}[\textbf{Causal Estimand Under Spatial Interference}]
\textit{Our target estimand is the RRT under spatial interference during the post-treatment period:
\begin{equation}\label{eq:rr-interference}
    \text{RRT}^{\text{SA-SC}} = \frac{\mathbb{E}[Y_{it}(1,\cdot) \mid i \in \mathcal{Z}_t, T_0 < t < T_0+\ell]}{\mathbb{E}[Y_{it}(0,0) \mid i \in \mathcal{Z}_t, T_0 < t < T_0+\ell]}.
\end{equation}
This captures the total effect of experiencing a heatwave  versus complete non-exposure to a heatwave.
\footnote{\scriptsize{Either the region is also exposed in this case, $Y_{it}(1,1)$ if not $Y_{it}(1,0)$.}}
The numerator represents observed outcomes for treated units embedded in treated/untreated neighborhoods; the denominator represents the counterfactual of no exposure from non-exposed donors.}
\end{definition}
$\text{RRT}^{\text{SA-SC}}$ directly estimates the effect of being in a heatwave region versus being completely outside any heatwave influence, which aligns with the physical reality of how extreme heat events affect geographic areas.

\begin{proposition}[\textbf{Identification Under Spatial Interference}]
\textit{Under Assumptions \ref{ass:overlap}--\ref{ass:sutnva}, the causal estimand $\text{RR}^{\text{SA-SC}}$ can be identified from the observed data and the post-treatment counterfactual is}
$$\widehat{Y}_{it}(0,0) = \sum_{j \in \mathcal{N}^t_0,\text{buf}}{\widehat{w}_j^i Y_{jt}}\; , \quad t > T_0\;.$$
\end{proposition}
\noindent \textit{Proof}: See Supplementary Materials.

This approach of identification under partial interference with SUTNVA and contiguous treatment acknowledges both the spatial spillovers and the regional nature of heatwaves. By incorporating the constraint that heatwaves affect contiguous regions of at least radius $s_0$, we provide a framework that is both meteorologically realistic and methodologically tractable.

The assumptions required for identification (Assumptions \ref{ass:overlap}--\ref{ass:noanticipation}) remain largely the same, with the modification that weights interpolation now requires that treated units embedded in heatwave regions can be approximated by a weighted combination of genuinely unexposed units. 
This is more restrictive in practice as the donor pool must be at least $s_0+1$ degrees away from any treated unit, substantially reducing its size, but ensures that our counterfactual estimates reflect true non-exposure rather than partial or spillover exposure.

Table \ref{tab:methods_comparison} provides a comprehensive comparison of methods for estimating heatwave effects, organized into two main categories: traditional statistical methods presented in Section \ref{sec:background} and the proposed spatial causal inference methods presented in this Section. For each approach, the table presents the key identifying assumptions required and the corresponding causal estimand targeted.
Having established identification under spatial interference, we now operationalize these ideas through a Bayesian framework that naturally incorporates spatial regularization. 
%The weights are estimated following \ref{eq:weights}, with the consideration of geographical distance across units. 
%\falco{The weights, are ... and, as we will see below, are identified...}

\input{Tables/methods_comparison}

\subsection{\textbf{Heatwave Effects Estimation: Spatially Augmented Bayesian Synthetic Controls}}

Under the identification framework developed in the previous section, the numerator in equation \eqref{eq:rr-interference} is equal to the expected value of the observed outcome of the treated units; thus, we only need to impute the missing potential outcome: the outcome for treated units under control assignment, that appears at the denominator of equation \eqref{eq:rr-interference}. 
Hence, we introduce a Bayesian estimator that implements the spatially-augmented synthetic control---SA-SC. This approach provides uncertainty quantification for counterfactual outcomes while incorporating spatial regularization through distance-dependent priors on donor weights.
For treated unit $i \in \mathcal{Z}_t$ experiencing a heatwave at $T_0$, let $\mathcal{O}_i = (\bm{Y}_{i,\mathcal{T}^-}, \mathbf{X}_i, \mathbf{Y}_{\mathcal{N}_0^{*t}})$ denote the observed data, where $\bm{Y}_{i,\mathcal{T}^-}$ is the pre-treatment outcome for $t \in \mathcal{T}^-$ where $\mathcal{T}^- = \{t: t \leq T_0\}$, $\mathbf{X}_i$ are covariates, and $\mathbf{Y}_{\mathcal{N}_0^{*t}}$ contains all donor unit outcomes. 
The posterior predictive distribution for the missing counterfactual outcomes $Y_{i,\mathcal{T}^+}(0,0)$ is:
$$
\pi(Y_{i,\mathcal{T}^+}(0,0)\mid \mathcal O_i)
 = \int_{\Theta}
   \pi(\bm{\theta} \mid \mathcal O_i)
   \prod_{t \in \mathcal{T}^+}
   \pi(Y_{it}(0,0)\mid \mathcal O_i,\bm{\theta})\,d\bm{\theta},
$$
where $\bm{\theta}$ denotes the full parameter vector. This formulation encodes Assumption~\ref{ass:stablepattern} (Stable Pattern) through the conditional independence structure. 
We model the pre-treatment outcomes as:
$$\bm{Y}_{i,\mathcal{T}^-}(0,0) \mid \bm\theta \sim \text{N}\left(\bm{Y}_{\mathcal{N}_0^{*t},\mathcal{T}^-} \omega^i, \sigma_i^2 I_{\mathcal{T}^-}\right)
$$
where $\bm{Y}_{\mathcal{N}_0^{*t},\mathcal{T}^-}$ is the $T_0 \times |\mathcal{N}_0^{*t}|$ matrix of donor pre-treatment outcomes, and $\omega^i \in \Delta^{|\mathcal{N}_0^{*t}|-1}$ are the synthetic control weights, estimated according \ref{eq:weights}. 
To incorporate spatial structure while maintaining the simplex constraint, we parameterize weights via distance-regularized logits. Let $\eta^i = (\eta_j^i)_{j \in \mathcal{N}_0^{*t}}$ be latent coefficients with:
$$
\omega_j^i = \frac{\exp(\eta_j^i - \bar{\eta}^i)}{\sum_{k \in \mathcal{N}_0^{*t}} \exp(\eta_k^i - \bar{\eta}^i)}, \quad \bar{\eta}^i = \frac{1}{|\mathcal{N}_0^{*t}|} \sum_{j \in \mathcal{N}_0^{*t}} \eta_j^i
$$
which guarantees $\omega_j^i > 0$ and $\sum_j \omega_j^i = 1$ while improving identifiability by centering $\eta^i$ \citep{atchison1980logistic}.
The spatial prior penalizes distant donors:
$$\eta_j^i \sim \text{N}(-\varsigma D_{ij}, \tau_\eta^2)
$$
where $D_{ij}$ is the geographic distance (in km) between unit centroids, $\varsigma > 0$ controls spatial decay, and $\sigma_\eta$ governs variability. This ensures nearby donors receive higher prior weight while maintaining differentiability.
Weakly-informative hyperpriors complete the hierarchy: $\sigma_i \sim \text{N}^+(0, 1), \quad \sigma_\eta \sim \text{N}^+(0, 1), \quad \varsigma \sim \text{N}^+(0, 1)$ where $\text{N}^+$ denotes a half-normal distribution.
We estimate parameters using Hamiltonian Monte Carlo. 
Model adequacy is assessed through posterior predictive checks on pre-treatment fit. The posterior mean counterfactual is:
$\bm{\hat{Y}}_{i,\mathcal{T}^+}(0,0) = \sum_{j \in \mathcal{N}_0^{*t}} \bar{\omega}_j^i Y_{j,\mathcal{T}^+}$
where $\bar{\omega}_j^i$ is the posterior mean weight. 
The aggregate relative risk is:
$$\widehat{\text{RRT}}^{\text{SA-SC}} = \frac{1}{|\mathcal{Z}_t|} \sum_{i \in \mathcal{Z}_t} \frac{\sum_{t \in \mathcal{T}^+} Y_{it}}{\sum_{t \in \mathcal{T}^+} \hat{Y}_{it}(0,0)}
$$
with uncertainty quantified through posterior samples. Unlike standard Bayesian SC with non-informed normal priors over space \citep{martinez2022bayesian}, our spatial regularization directly encodes the meteorological reality that proximate areas share weather patterns and vulnerability factors. This spatial structure yields more coherent counterfactuals while preserving the transparency of causal identification.

\section{Empirically-based Monte Carlo Simulations}\label{sec:simulation}

In the following, we validate the performance of the proposed causal methods through empirical Monte Carlo simulations where covariates and exposure come from real-world data while outcomes are generated with known causal effects so that we can access the \textit{ground truth} to evaluate the performance of the proposed models. 
The next sections describe the data, the simulation design and the results. 

\vspace{-0.25cm}
\subsection{\textbf{Data Setup and Data Sources}}

We focused on counties in Northeastern United States: Connecticut, Maine, Massachusetts, New Hampshire, New Jersey, New York, Pennsylvania, Rhode Island and Vermont. 
Meteorological exposures were derived from the GridMET dataset \citep{abatzoglou2013development}, which provides gridded daily estimates of temperature, precipitation, humidity, and other variables across the contiguous United States at 4 km spatial resolution. 
We computed population-weighted daily averages for each county. The heat index was calculated following \citep{anderson2013methods} using daily maximum temperature and precipitation. 
Additional covariates at a yearly resolution including racial composition, poverty rate, median household income, median home value, and educational attainment, were linked from the American Community Survey.

%\subsection{\textbf{Validation with Simulated Outcomes}}\label{sec:simulation}

\subsection{\textbf{Simulation Design}}

In our simulations, we generate the outcome to have full control over the ground truth and be able to evaluate the performance of the proposed causal inference methods.
We encode spatial dependence through a sparse, row-stochastic weight matrix $\bm {W_{geo}}$ constructed using a $k$-nearest-neighbor kernel with $k=4$. 
The weights are proportional to $\exp(-D_{ij}/c)$ where $D_{ij}$ is the geographic distance between county centroids and $c=0.5c_0$, with $c_0$ defined as the median distance among each county's $k$ nearest neighbors.
The data generating process for potential outcomes under no treatment takes the form: $Y_{it}(0, 0)=\alpha_i+\delta_t+ f_t^\top B_{i}+u_{it}+\varepsilon_{it}$.
This structure captures multiple sources of variation. 
The term $\alpha_i$ represents spatially-smoothed unit-specific effects that follow a simultaneous autoregressive (SAR) structure: $\bm{\alpha}=(I-\varpi_\alpha  \bm W_{geo})^{-1}\nu_\alpha$ where $\nu_\alpha\sim \text{N}(0,\sigma_\alpha^2 I)$. 
These unit effects capture persistent differences across counties in baseline mortality rates, health infrastructure, and demographic composition.
\footnote{ \scriptsize We used mortality rates because data on mortality are publicly available through CDC Wonder and our synthetic outcome can be ancored on their real distribution \citep{}, while hospitalization are not publicly available so are used only in the real-world application. 
However, any outcome can be used.} 
The common time effects $\delta_t$ follow an AR(1) process and represent seasonal mortality patterns and secular trends. 
The interactive term $f_t^\top B_i$ captures time-varying heterogeneity through a 2-dimensional AR(1) latent factor $f_t$ combined with spatially-smoothed county-specific loadings $B_i$ (constructed analogously to $\alpha_i$). 
This allows different counties to respond differently to common shocks. 
The idiosyncratic error $u_{it}$ propagates spatially according to $u_{it}=\rho_u \bm W_{geo} u_{it-1}+\epsilon_t$, while $\varepsilon_{it}$ represents pure measurement noise.
%We standardize daily mean heat index $H_{it}$ using the pre-period mean and standard deviation to obtain $\tilde{H}_{it}$. 
For treated units after the heatwave onset, we generate treatment effects as $\tau_{it} = \tau_0(\exp(\kappa \tilde{H}_{it})-1)$, where the exponential form ensures that effects are small for mild heat deviations but accelerate at extreme values, consistent with epidemiological evidence of threshold effects.
To capture realistic spillover patterns, we allow treatment effects to propagate to neighboring untreated units. 
Specifically, an untreated unit $j$ located $s$ degrees of separation from a treated unit $i$ receives a spillover effect $\psi^{s}_{jt} = \chi^s \bm W_{adj} \tau_{it}$. 
We consider spillovers at one and two degrees of separation ($s=\{1,2\}$), with decay parameters $\chi = (0.70, 0.40)$ reflecting the empirical regularity that spillover effects diminish with distance. 
This generates three types of observed outcomes: treated units with treated neighbors experience $Y_{it} = Y_{it}(0, 0) \times \tau_{it}$; untreated units with treated neighbors experience $Y_{it} = Y_{it}(0, 0) \times \psi^s_{it}$; and units far from any heatwave experience baseline outcomes $Y_{it} = Y_{it}(0, 0)$.
\footnote{\scriptsize As observed in Section \ref{sec:method}, the configuration $Y_{it}(1,0)$ would require an isolated heatwave affecting a single unit, which is very rare for the spatial scale of our analysis (e.g., county-level). Thus, this configuration is excluded from our setup.} 
We calibrate the simulated pre-treatment levels to match observed county-specific mortality rates through additive shifts that preserve the treatment structure.
We generated 100 independent datasets, each based on real spatial and temporal characteristics. We use as set of units the counties in the Northeastern of U.S. (ME, MA, CT, RI, NH, NY, NJ, PE, VT). 
For each replication, we randomly sampled a Northeastern U.S. county and a year between 2000 and 2016, using the others as controls.
We check that our simulated log-mortality rates aligned with the empirical distribution observed in those counties during that period.

Our simulation design crossed two factors---(i) presence or absence of spatial dependence, and (ii) presence or absence of spillover effects---yielding four distinct scenarios. 
We also examined robustness to donor pool misspecification by comparing performance when the pool is correctly specified versus when it includes contaminated units (results for this scenario are reported in the Appendix). 
All simulation parameters were calibrated to reproduce realistic effect magnitudes (further details are provided in the Appendix).

\subsection{\textbf{Simulation Results}}

\input{Tables/sim_main}
Table \ref{tab:sim_res} depicts the results for absolute bias (\textit{abs. bias}), root mean squared error in the pre-treatment period (\textit{RMSE pre}), root mean squared error in the post-treatment period (\textit{RMSE post}), coverage of credible intervals (\textit{coverage}, and average length of the credible intervals (\textit{Avg. CI}). 
Details and formulas for these metrics can be found in the Appendix.

The SA-SC systematically outperforms standard SC across all performance metrics. SA-SC achieves lower absolute bias and smaller root mean squared errors in both the pre-treatment and post-treatment periods. 
The credible intervals are slightly shorter while maintaining better calibration, as evidenced by higher coverage probabilities. 
Importantly, these gains persist even when we introduce spatial dependence or spillover effects into the data generating process, suggesting that explicitly modeling spatial dependence helps the method recover latent structure and absorb cross-unit contamination.

While both methods exhibit coverage rates modestly below the nominal 95\% level, SA-SC consistently achieves higher coverage, suggesting that the spatial regularization provides more honest uncertainty quantification. 
We also examined sensitivity to spillover when assumption \ref{ass:sutnva} is not holding---that is, using all controls in donor pool, even those exposed to spillover. 
Appendix Table \ref{tab:sim_res_sutva} reports the results. 
Spatially–Augmented Synthetic Control (SA–SC) improves estimation accuracy when outcomes display spatial dependence but no spillovers occur, yielding lower bias, smaller RMSE, and better coverage than the standard SC. 
However, when spillover effects are present and not explicitly modeled, SA–SC performs worse, as it tends to overweight nearby control units that are themselves partially affected by treatment. 
In scenarios combining both spatial dependence and spillovers, the two estimators achieve similar bias and RMSE, though SA–SC maintains slightly poorer coverage. 
Overall, spatial augmentation offers clear gains when spatial structure drives confounding, but its advantage diminishes if spillover contamination is ignored.

\section{\textbf{Application to Hospitalization Outcomes from Medicare}}\label{sec:application}

In this Section, we apply the proposed methods to Medicare hospitalization data covering 13,753,273 beneficiaries residing in Northeastern U.S. counties from 2000 to 2019. The next sections describe the data (outcome and exposure) and results.

\subsection{\textbf{Outcome Definition}}

We next apply the proposed causal inference framework to estimate effects on heat-related hospitalization outcomes among older adults in the United States. 
Hospitalization data were obtained from the the Centers for Medicare \& Medicaid Services (CMS), the U.S. federal agency that maintains comprehensive administrative claims records for all Medicare beneficiaries nationwide.
Medicare provides health insurance to nearly all Americans aged 65 and older.
This population is particularly susceptible to adverse health effects from extreme heat \citep{bobb_cause-specific_2014}.

We aggregated individual inpatient claims to the county-day level and categorized according to the principal diagnosis recorded at discharge. 
Diagnoses were identified using International Classification of Diseases (ICD) codes.\footnote{\scriptsize ICD-9-CM was used before the national transition to ICD-10-CM (pre-2015), after which ICD-10-CM (post-2015) codes apply.}
We focused on hospitalization rates for conditions previously linked to heat stress among older adults \citep{bobb_cause-specific_2014}.%bargagli2025extreme
The diagnostic groups were: \textit{heat-related illness} (ICD-9-CM 992.x; ICD-10-CM T67.x), \textit{fluid and electrolyte disorders} (276.x; E86.x), \textit{acute kidney failure} (584.x; N17.x), \textit{urinary tract infection} (599.0; N39.0), and \textit{septicemia and severe systemic infections} (038.x; A41.x). 
Rural counties and counties with population below 200,000 were excluded from estimation because heat-related hospitalizations in those areas are rare, yielding sparse outcome data that violate the approximate normality assumptions required for standard synthetic control inference.
For each urban county and calendar day, we counted hospitalizations within these groups, obtained the number of Medicare fee-for-service beneficiaries residing in that county in the corresponding year, and computed daily hospitalization rates per 100,000 beneficiaries. 

The final analytic dataset included 13,753,273 beneficiaries in 217 counties that could serve as controls and 8,022,444 beneficiaries in 43 urban counties that were estimated across the Northeastern United States over the 2000–2019 period\footnote{\scriptsize Beneficiaries could enter or exit Medicare during the study, and some beneficiaries moved between urban and rural counties.}. Further details on the data and pre-analysis processing are provided in the Appendix.

\subsection{\textbf{Heatwave Definition and Treatment Assignment}}

Heatwaves were defined as periods of two or more consecutive days on which the county-specific daily mean heat index exceeded the 95th percentile of the county's yearly distribution.
To ensure clean causal identification and avoid overlapping exposure periods, we restricted our analysis to the first heatwave of each season for each county. 
This restriction simplifies causal interpretation by eliminating concerns about temporal carry-over effects from previous heatwaves within the same season.
While our method can be extended to analyze any heatwave event, researchers applying it to subsequent heatwaves should carefully consider potential biases from temporal carry-over effects \citep{papadogeorgou2022causal}.\footnote{\scriptsize Specifically, one should ensure that: (i) the pre-treatment period is uncontaminated by prior heatwave exposures, and (ii) a sufficiently long washout period separates consecutive heatwaves to allow for full dissipation of any lingering effects. The required washout period may vary depending on the specific health outcome under study and local environmental conditions.}

\subsection{\textbf{Results}}

We applied both the SC and SA-SC methods under the SUTNVA assumption, which allows interference within a county's immediate spatial neighborhood and permits potential outcomes to depend on the treatment status of nearby counties. 
Under this assumption, causal effects are identified by comparing treated counties to donor counties that lie outside the defined spillover radius. 
For each urban county and each study year between 2000 and 2019, we estimated the effect of the first heatwave of the season.

\begin{figure}
    \centering
    \includegraphics[width=.85\linewidth]{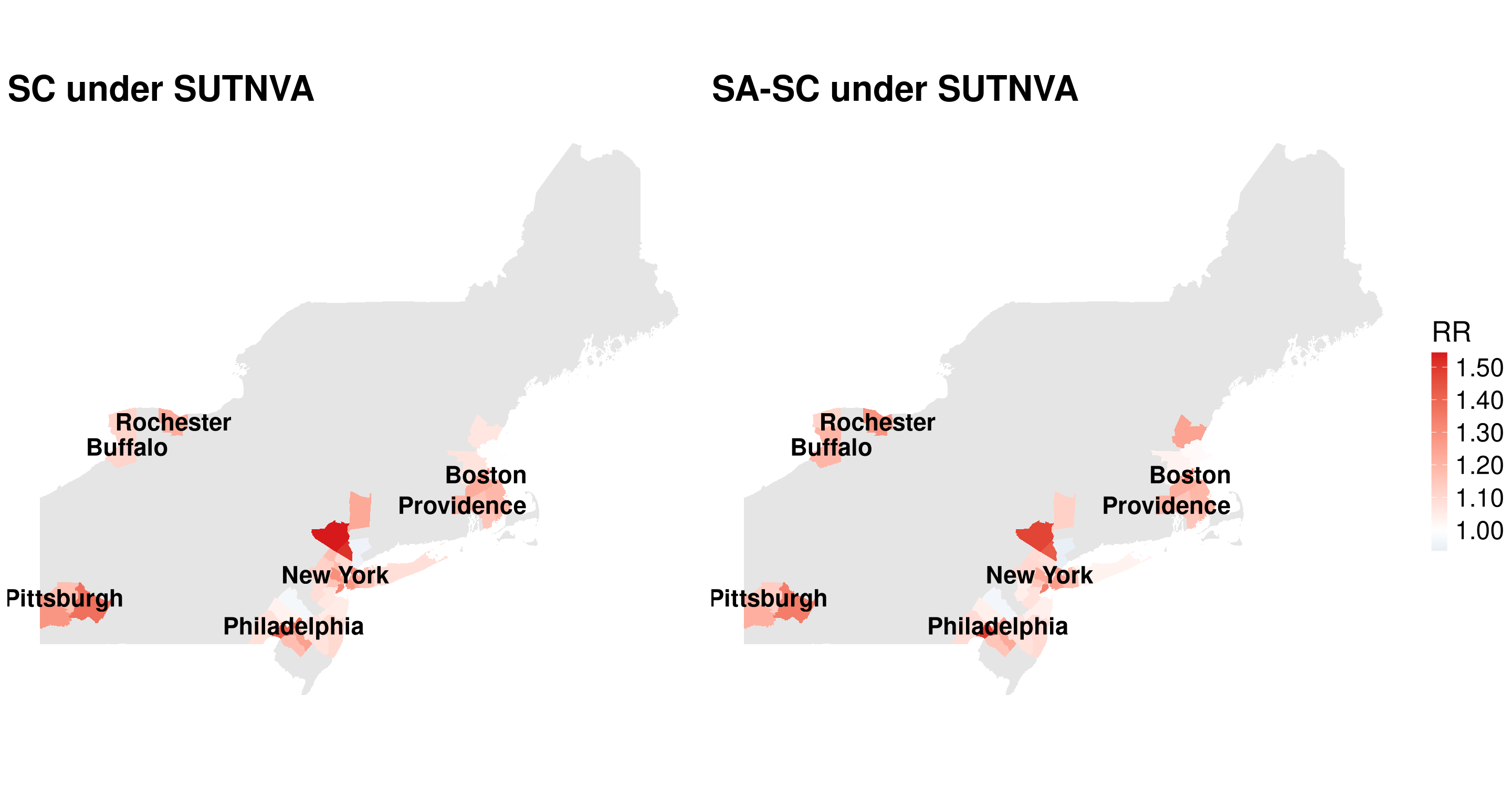}
    \caption{Posterior mean relative risk of heat-related hospitalizations during heatwaves versus non-heatwave days across urban Northeastern U.S. counties, 2000-2019. Heatwaves are defined as periods of at least two consecutive days exceeding the 95th percentile of county-specific annual heat index values. Estimates correspond to first-season heatwaves with donor pools restricted to non-exposed counties outside treated neighborhoods. Left panel: standard synthetic control (SC). Right panel: spatially-augmented synthetic control (SA-SC). Red indicates increased hospitalization risk (RR $>$ 1) during heatwaves; grey indicates counties excluded from estimation.}
    \label{fig:mean-map}

\end{figure}

Figure \ref{fig:mean-map} displays the posterior mean relative risk of heat-related hospitalizations during heatwaves, averaged across all study years from 2000 to 2019. 
Both models under the SUTNVA assumption produce broadly similar estimates. 
Most urban counties show elevated mean hospitalization risk, with posterior mean relative risks ranging from 0.95 to 1.55. 
The highest risks appear in highly populated urban areas and adjacent counties around the New York City metropolitan area.  
The lowest risks occur in less densely populated urban counties around Boston, Providence, and Philadelphia.
County-level credible intervals include 1 (see Figure \ref{fig:plot-counties}). 
Since our primary estimand is defined at the county–year level under partial interference, these intervals describe local uncertainty for each first-season event rather than a regionally pooled effect.

To obtain a summary measure that is comparable to multi-community estimates in the existing literature, we conducted a random effects meta analysis on the log scale using the metafor package \citep{viechtbauer2010conducting}. 
Pooling across all first-season heatwaves from 2000 to 2019, the standard SC model produced a pooled relative risk of 1.10 with a 95 percent interval from 1.07 to 1.14. The SA–SC model yielded a pooled relative risk of 1.11 with a 95 percent interval from 1.08 to 1.15. 
These pooled estimates indicate a regionally averaged increase of about 10-11 percent in heat-related hospitalization rates during first-season heatwaves among older adults in urban Northeastern counties, consistent with previous findings reported in the literature \citep{bobb_cause-specific_2014} providing validation while extending these findings to account for spillovers.

Results were similar under alternative heatwave definitions and showed little temporal variation in county–year effects. 
The Appendix provides further details on model implementation, including the construction of donor pools, and reports additional sensitivity analyses using different heatwave thresholds, alternative temporal specifications, and posterior intervals for all counties.

\section{Discussion}\label{sec:discussion} 

This paper develops a causal inference framework for estimating heatwave effects that addresses fundamental limitations in existing epidemiological approaches. 
We contribute: 
(i) systematic critical analysis of widely-used methods (DLNMs, Bayesian hierarchical models, case-crossover designs), explicating their implicit causal assumptions and demonstrating where these might break down under realistic scenarios in the heatwave context; 
(ii) a transparent causal identification framework using synthetic controls for heatwave effect estimation and relaxing SUTVA through partial interference, introducing spatially-augmented synthetic control methods explicitly modeling spatial dependence and spillovers; and 
(iii) validation through empirically-based Monte Carlo simulations and application to Medicare hospitalizations in urban Northeastern U.S. counties (2000-2019), demonstrating that our proposed methods perform well in terms of root mean squared error and posterior interval coverage with SA-SC outperforming traditional SC methods under spatial dependencies and spillover effects. 
Several features of our framework warrant extensive discussion and invite debate.

\subsection{\textbf{Design-informed Statistical Modeling Causal Inference Framework for Informing Policies and Actions}}

The longstanding debate in statistics---especially salient in official statistics and survey methodology---has traditionally centered on \textit{design-based} versus \textit{model-based} inference--from \cite{sarndal1978design} to \cite{newzhang}, among others. 
In the classic setting, design-based approaches anchor inference in the randomization provided by probability sampling, while model-based approaches rely on statistical models to extrapolate beyond observed data, sometimes with limited transparency about their assumptions. 
This tension, and eventually the interplay between the two approaches \citep[e.g.,][]{yu2018design}, have shaped methodological progress in survey sampling and provided a rich foundation for statistical reasoning about representativeness and uncertainty.

We believe that our framework exemplifies a new fundamental tension in applied statistics, which plays out between \textit{statistical modeling} capturing complex associations versus \textit{causal design} targeting interpretable estimands under explicit assumptions. 
Similar to the traditional design-based inference, the causal design approach relies on the concept of randomization; however, the source of uncertainty comes from the (quasi-)randomization of the treatment rather than the sampling. Different from the traditional statistical modeling approaches, that provide valuable associational evidence but seldom make their identifying assumptions explicit or verify their plausibility, the ``new'' design-based causal frameworks demonstrate that establishing the ``\textit{rules of the game}''---that is, clearly defining what identifying assumptions are needed for causal claims and what causal quantities can be estimated from the data---is essential for answering relevant research questions in a causal perspective.

In this work, we demonstrate that the interplay between rigorous causal design and statistical modeling lays the foundation for transparent and credible causal evidence that can inform policy decisions.
This transparency is particularly critical when leveraging quasi-experimental variation from natural phenomena like heatwaves, where treatment assignment involves both exogenous meteorological processes but potentially endogenous responses. 
Specifically, explicitly stating causal assumptions and relaxing restrictive independence conditions in favor of more realistic spatial interference models provides a framework that balances causal validity with statistical efficiency.

In this perspective, the advantages of a \textit{design-informed statistical modeling} causal inference framework over a purely \textit{modeling-oriented} approach are not merely statistical but they can also inform resource allocation, intervention design, and policy priorities under constraints. 
The distinction between causal and associational estimates matters precisely because interventions are costly and geographically bounded.
We argue that a causal inference framework in heatwave epidemiology can generate policy-relevant evidence by answering three fundamental questions to optimize social benefit.
First, \textit{where to target}. Unbiased effect estimates enable evidence-based prioritization: large causal effects on certain areas or heat-related illness justify targeted intervention policies (e.g., localized cooling center interventions).
Second, \textit{how far to reach}. The spillover radius directly informs intervention scope: if health effects extend, say, 50 kilometers from the most vulnerable areas, cooling center placement strategies must cover correspondingly large geographic areas and require regional coordination rather than jurisdiction-by-jurisdiction responses.
Third, \textit{who to target}. Understanding effect heterogeneity across vulnerable populations (by, e.g., age, income, comorbidities, air conditioning access) informs who to protect first under resource constraints. 

Our contribution focuses primarily on framework and methodological development rather than exhaustively demonstrating all three policy mechanisms. 
However, the framework  we provide lays the foundations for future research that can fully operationalize these mechanisms with the ultimate goal of informing heat mitigation policies. 

\subsection{\textbf{Spatial Methods in Causal Inference: Bridging Disciplines}}

A central methodological contribution is explicitly incorporating spatial structure into causal inference frameworks.
By relaxing SUTVA and accounting for spatial interference and dependence constructing spatially-buffered donor pools with distance-regularized priors, we provide a framework that is both meteorologically realistic and methodologically tractable.
This connects spatial statistics, which has long emphasized dependence structures, with causal inference's potential outcomes framework, bridging disciplines that have often developed largely independently.

Our framework synthesizes methods from multiple disciplines: causal inference (potential outcomes, identification assumptions), spatial statistics (spatial dependence, spillover modeling), environmental epidemiology (exposure assessment, health outcomes), and climate science (extreme event definition).
Such interdisciplinary integration reflects that environmental health problems require methodological innovation at disciplinary intersections: neither purely spatial statistical approaches that ignore causal identification, nor purely causal approaches that ignore spatial structure, suffice.

The methods apply broadly beyond heatwaves to environmental and policy evaluations with spatially structured treatments: e.g., wildfire smoke exposure, flooding events, cooling center programs, heat warning systems, climate adaptation infrastructure, and emergency response protocols.
Each involves spatial contiguity and potential cross-boundary spillovers where standard no-interference assumptions would misattribute effects.
The framework extends naturally to network structures beyond geography such as social networks, trade networks, infrastructure networks—through appropriate network-based priors \citep{forastiere2021identification}.
\subsection{\textbf{Limitations, Methodological Choices, and Future Directions}}

Some design choices warrant discussion.
Our distance-based prior encodes that nearby donors provide more credible counterfactuals, but our spillover buffer represents a fundamental trade-off: larger buffers reduce contamination bias but shrink donor pools and may worsen synthetic control fit. 
What constitutes the ``optimal'' buffer varies by outcome type, population density, and local infrastructure.
Similarly, our 60-day pre-treatment period balances learning stable patterns against temporal instability from adaptation, but the optimal length remains context-dependent.
These choices involve judgment and domain knowledge, making them natural points for future investigation and refinement.

A further limitation arises from our restriction to urban counties. 
Rural counties were excluded because heat-related hospitalizations in those areas are rare, leading to outcome distributions that violate the approximate normality assumption underpinning standard synthetic control estimators. 
In such sparse-outcome settings, Gaussian error models yield unstable or undefined likelihoods, making reliable estimation infeasible.
Future research should account for this shortcoming to produce robust evidence even in settings where the outcome is rare.

Beyond design choices, several substantive limitations suggest directions for future research.
Our estimand targets average effects across treated units, but substantial vulnerability heterogeneity means average effects may obscure distributional considerations.
Future work could estimate conditional average treatment effects (CATE) under network interference, leveraging the Bayesian framework for hierarchical uncertainty propagation.
We focus on contemporaneous effects, but health impacts exhibit complex temporal dynamics—harvesting effects, adaptation, cumulative damage suggesting extensions to dynamic causal effects through time-varying parameters or distributed lag structures within SC frameworks.
Our binary heatwave definition facilitates clean interpretation but discards intensity variation; continuous treatment methods or dose-response functions could accommodate this, though extending to spatial interference settings remains underdeveloped.
Focusing on first-season heatwaves enables clean identification but limits generalizability to frequent, repeated exposures; methods for multiple treatment episodes with spatiotemporal spillovers would enable longer time series analysis.
Machine learning integration could also improve donor selection and flexible outcome modeling, though maintaining interpretability and uncertainty quantification in Bayesian SC remains challenging.

%\subsection{Conclusion}

As climate extremes intensify, rigorous causal inference in environmental health becomes increasingly critical for designing effective adaptation policies.
Our contribution is providing a transparent framework that explicitly addresses spatial dependencies and spillovers in heatwave epidemiology while maintaining clear causal estimands allowing to estimate not merely associations but the effects defined heatwave events.
We hope this work stimulates discussion about further integrating causal thinking into environmental epidemiology: acknowledging methodological trade-offs, debating design and modeling choices, and ultimately improving the evidence base for protecting vulnerable populations from climate-related health threats.

%\section*{Funding and Acknowledgements}

%This work was funded by grants...

%\vspace{1em} 

% \noindent We benefited from helpful comments and suggestions from ..., and participants at the Harvard Causal Inference workshop, at the NSAPH Biostatistics seminar, and at the ...

{
\bibliographystyle{abbrvnat}
\bibliography{bibtex}
}

%%%%%%%%%%%%%%
\clearpage
\appendix
\section*{Supplementary Material for ``Estimating the Effects of Heatwaves on Health: A Causal Inference Framework'', Giulio Grossi, Leo Vanciu, Veronica Ballerini, Danielle Braun, Falco J.\ Bargagli-Stoffi}
\renewcommand{\thesubsection}{\Alph{subsection}}
\renewcommand\thefigure{\thesubsection.\arabic{figure}}   
\renewcommand\thetable{\thesubsection.\arabic{table}}  
\counterwithin{figure}{section}
\counterwithin{table}{section}
\numberwithin{equation}{section}
 \pagenumbering{arabic}
    \setcounter{page}{1}

\input{Sections/appendix}

%TC:endignore

\end{document}

%% file: Tables/methods_comparison.tex
\begin{table*}%[ht!]
\centering
\resizebox{\textwidth}{!}{%
\begin{tabular}{@{}lp{5.5cm}l@{}}
\toprule
\textbf{Method} & \textbf{Key Assumptions} & \textbf{Causal Estimand} \\
\midrule
\multicolumn{3}{l}{\textit{Traditional Statistical Methods}} \\
\midrule
Multi-Community/GEE  (Section \ref{ss:gee})  & 
(i) SUTVA; (ii) Conditional Ignorability;  (iii) Treatment Overlap & 
$\text{RR}^{\text{GEE}} = \mathbb{E}[Y(1)\mid C_i] / \mathbb{E}[Y(0)\mid C_i]$ \\
\addlinespace
Hierarchical Bayesian  (Section \ref{ss:bayes}) & 
(i) SUTVA; (ii) Conditional Ignorability; (iii) Treatment Overlap & 
$\text{RR}_i^{\text{HB}} = \mathbb{E}[Y(1)\mid \alpha_i] / \mathbb{E}[Y(0)\mid \alpha_i]$ \\
\addlinespace
DLNM (Section \ref{ss:dlnm}) & 
(i) SUTVA; (ii) Conditional Ignorability; (iii) Treatment Overlap & 
$\text{RR}^{\text{DLNM}} = \mathbb{E}[Y(1)\mid\mathbf{X}_{it}] / \mathbb{E}[Y(0)\mid\mathbf{X}_{it}]$ \\
\addlinespace
Matching Design  (Section \ref{ss:cc}) & 
(i) SUTVA; (ii) Temporal Exchangeability; (iii) Treatment Overlap; (iv) No Carry-over; (iv) Stable Composition & 
$\text{RR}_i^{\text{CC}} = \mathbb{E}[Y_{it}(1)] / \mathbb{E}[Y_{it'}(0)]$ \\
\midrule
\multicolumn{3}{l}{\textit{Proposed Spatial Causal Inference Methods}} \\
\midrule
Synthetic Control  (Section \ref{ss:synth}) & 
(i) SUTVA; (ii)  Weights Interpolation; (iii) Stable Pattern; (iv) No Anticipation; (v) Idiosyncratic Shocks & 
$\text{RRT}^{\text{SC}} = \mathbb{E}[Y(1)\mid Z_t, T_0 < t < T_0+\ell] / \mathbb{E}[Y(0)\mid Z_t, T_0 < t < T_0+\ell]$ \\
\addlinespace
Spatial-SC  (Section \ref{ss:sasynth}) & 
(i) \textbf{SUTNVA} (Allows Spillovers); (ii) Weights Interpolation; (iii) Stable Pattern; (iv) No Anticipation; (v) Idiosyncratic Shocks & 
$\text{RRT}^{\text{SA-SC}} = \mathbb{E}[Y(1,\cdot)\mid Z_t, T_0 < t < T_0+\ell] / \mathbb{E}[Y(0,0)\mid Z_t, T_0 < t < T_0+\ell]$ \\
\bottomrule
\end{tabular}%
}
\caption{Comparison of Methods for Estimating Heatwave Effects}\label{tab:methods_comparison}
\label{tab:methods_comparison}
\vspace{-0.5cm}
\end{table*}

%% file: Tables/sim_main.tex
\begin{table}
    \centering
    \resizebox{\linewidth}{!}{
\begin{tabular}{llrrrrr}
\toprule
  & Method & Abs avg. Bias & RMSE pre & RMSPE post & Coverage prob. & Avg. CI length\\
\midrule
No Spatial depend. - No spillover & SC & 0.820 & 1.311 & 1.455 & 0.936 & 3.744\\ \rowcolor{lightgray}
 & SA-SC & 0.718 & 1.210 & 1.377 & 0.948 & 3.471\\
No Spatial depend. - Spillover & SC & 0.829 & 1.313 & 1.443 & 0.937 & 3.753\\ \rowcolor{lightgray}
 & SA-SC & 0.718 & 1.085 & 1.329 & 0.948 & 3.471\\
Spatial depend. - No Spillover & SC & 0.830 & 1.334 & 1.474 & 0.931 & 3.770\\ \rowcolor{lightgray}
 & SA-SC & 0.694 & 1.093 & 1.330 & 0.939 & 3.383\\
Spatial depend. - Spillover & SC & 0.833 & 1.295 & 1.490 & 0.933 & 3.769\\ \rowcolor{lightgray}
& SA-SC & 0.694 & 1.106 & 1.361 & 0.939 & 3.383\\
\bottomrule
\end{tabular}
    }
    \vspace{-0.1cm}
    \caption{Monte Carlo performance across 100 replications for causal inference methods.  SC refers to \textit{synthetic control} while SA-SC refers to \textit{spatially augmented synthetic control}.}
    \label{tab:sim_res} 
    \vspace{-0.5cm}
\end{table}

%% file: Sections/appendix.tex
\vspace{1cm}

\section{Notation}

Table \ref{tab:notation} below summarizes the full notation used throughout the paper to assist readers in following the methodological development and empirical analyses.

\input{Tables/notation_table}

\pagebreak

\section{Causal Identification}

\subsection*{\textbf{Causal Identification of State-of-the-art Models}}

\subsubsection*{\textbf{Regression-Based Multi-Community Approach}}

Recall the model:
\begin{equation}
\log(\mathbb{E}[Y_{it}]) = \alpha_{\textbf{C}} + \beta Z_{it} + \mathbf{X}_{it}'\boldsymbol{\gamma}.
\end{equation}
Recall the assumptions:
\begin{enumerate}
\item[A1. ] SUTVA: $Y_{it}(\mathbf{z}) = Y_{it}(z_i)$ (no interference) and $Y_{it}(z_i) = Y_{it}$ when $Z_{it} = z_i$ (consistency);
\item[A2. ] Conditional ignorability: $(Y_{it}(1), Y_{it}(0)) \perp\!\!\!\perp Z_{it} \mid C_i, \mathbf{X}_{it}$;
\item[A3. ] Treatment Overlap: $0 < P(Z_{it}=1 \mid C_i, \mathbf{X}_{it}) < 1$.
\end{enumerate}
Then, one can identify from the data:
\begin{align}
\mathbb{E}[Y_{it}(1) \mid C_i, \mathbf{X}_{it}] &= \mathbb{E}[Y_{it}(1) \mid Z_{it}=1, C_i, \mathbf{X}_{it}]  
            \quad \because \text{ A2} \nonumber \\
            &= \mathbb{E}[Y_{it} \mid Z_{it}=1, C_i, \mathbf{X}_{it}]  
            \quad \because \text{ A1, A3} \nonumber \\
\mathbb{E}[Y_{it}(0) \mid C_i, \mathbf{X}_{it}] &= \mathbb{E}[Y_{it}(0) \mid Z_{it}=0, C_i, \mathbf{X}_{it}]  
            \quad \because \text{ A2} \nonumber \\
            &= \mathbb{E}[Y_{it} \mid Z_{it}=0, C_i, \mathbf{X}_{it}]  
            \quad \because \text{A1, A3} \nonumber
\end{align}
therefore:
\begin{align}
\text{RR}^{\text{GEE}} &= \frac{\mathbb{E}[Y_{it}(1) \mid C_i, \mathbf{X}_{it}]}
                    {\mathbb{E}[Y_{it}(0) \mid C_i, \mathbf{X}_{it}]} \nonumber \\
            &= \frac{\mathbb{E}[Y_{it} \mid Z_{it}=1, C_i, \mathbf{X}_{it}]}
                    {\mathbb{E}[Y_{it} \mid Z_{it}=0, C_i, \mathbf{X}_{it}]}  
            \quad \because \text{ Identification above} \nonumber \\
            &= \exp(\beta) 
            \quad \because \text{ Correct model specification.}
\end{align}
Effect homogeneity or correctly specified heterogeneity model assumptions are needed to pool effects across communities.

We note that here and below identification is provided for conditional estimands  (i.e., conditioning on covariates, community, or unit-specific effects). These conditional estimands can be used to obtain unconditional (population-average) estimands via the law of iterated expectations.

\subsubsection*{\textbf{Hierarchical Bayesian Approach}}
Recall the model:
\begin{equation}
\begin{split}
Y_{it} &\sim \text{Poisson}(\zeta_{it}) \\
\log(\zeta_{it}) &= \alpha_i + \beta_i Z_{it} + v(\mathbf{X}_{it}; \boldsymbol{\gamma}_i), \quad
\beta_i \sim \mathcal{N}(\mu, \sigma^2).
\end{split}
\end{equation}
where $v(\cdot)$ represents smooth functions of confounders including temperature and temporal trends.
Recall the assumptions:
\begin{enumerate}
\item[B2.] Conditional ignorability: $(Y_{it}(1), Y_{it}(0)) \perp\!\!\!\perp Z_{it} \mid \alpha_i, \mathbf{X}_{it}$;
\item[B3.] Treatment Overlap: $0 < P(Z_{it}=1 \mid \alpha_i, \mathbf{X}_{it}) < 1$.
\end{enumerate}
Then, one can identify from the data:
\begin{align}
\mathbb{E}[Y_{it}(1) \mid \alpha_i, \mathbf{X}_{it}] &= \mathbb{E}[Y_{it}(1) \mid Z_{it}=1, \alpha_i, \mathbf{X}_{it}]  
            \quad \because \text{ B2} \nonumber \\
            &= \mathbb{E}[Y_{it} \mid Z_{it}=1, \alpha_i, \mathbf{X}_{it}]  
            \quad \because \text{ A1, B3} \nonumber \\
\mathbb{E}[Y_{it}(0) \mid \alpha_i, \mathbf{X}_{it}] &= \mathbb{E}[Y_{it}(0) \mid Z_{it}=0, \alpha_i, \mathbf{X}_{it}]  
            \quad \because \text{ B2} \nonumber \\
            &= \mathbb{E}[Y_{it} \mid Z_{it}=0, \alpha_i, \mathbf{X}_{it}]  
            \quad \because \text{ A1, B3} \nonumber
\end{align}
therefore:
\begin{align}
\text{RR}_i^{\text{HB}} &= \frac{\mathbb{E}[Y_{it}(1) \mid \alpha_i, \mathbf{X}_{it}]}
                    {\mathbb{E}[Y_{it}(0) \mid \alpha_i, \mathbf{X}_{it}]} \nonumber \\
            &= \frac{\mathbb{E}[Y_{it} \mid Z_{it}=1, \alpha_i, \mathbf{X}_{it}]}
                    {\mathbb{E}[Y_{it} \mid Z_{it}=0, \alpha_i, \mathbf{X}_{it}]}  
            \quad \because \text{ Identification above} \nonumber \\
            &= \exp(\beta_i) 
            \quad \because \text{ Correct model specification.}
\end{align}
Model specification assumes that $\alpha_i$ and $v(\mathbf{X}_{it}; \boldsymbol{\gamma}_i)$ capture all confounders. For heat wave characteristics, \cite{anderson2011heat} extend this framework with $\beta_i \sim \mathcal{N}(\alpha_0 + \alpha_1 \mathbf{W}_i, \tau^2)$ where $\mathbf{W}_i$ includes intensity, duration, and timing. \cite{bobb2011bayesian} employ Bayesian model averaging over different specifications of $v(\cdot)$ to account for model uncertainty.

Furthermore, exchangeability assumption---that is, $P(\beta_i \mid \mathbf{W}_i) = P(\beta_j \mid \mathbf{W}_j)$ when $\mathbf{W}_i = \mathbf{W}_j$---is often invoked as it allows borrowing strength across communities for estimation.

\subsubsection*{\textbf{Distributed Lag Non-Linear Models (DLNM)}}
Recall the model:
\begin{equation}
\log(\mathbb{E}[Y_{it}]) = \alpha + h(H_{it}, \text{lag}) + \gamma Z_{it} + g(\text{t}) + \mathbf{X}_{it}'\boldsymbol{\beta}.
\end{equation}
Recall the assumptions:
\begin{enumerate}
\item[C2. ] Conditional ignorability: $(Y_{it}(1), Y_{it}(0)) \perp\!\!\!\perp Z_{it} \mid H_{it}, g(\text{t}), \mathbf{X}_{it}$;
\item[C3. ] Treatment Overlap: $0 < P(Z_{it}=1 \mid H_{it}, g(\text{t}), \mathbf{X}_{it}) < 1$.
\end{enumerate}
Then, one can identify from the data:
\begin{align}
\mathbb{E}[Y_{it}(1) \mid H_{it}, g(\text{t}), \mathbf{X}_{it}] &= \mathbb{E}[Y_{it}(1) \mid Z_{it}=1, H_{it}, g(\text{t}), \mathbf{X}_{it}]  
            \quad \because \text{ C2} \nonumber \\
            &= \mathbb{E}[Y_{it} \mid Z_{it}=1, H_{it}, g(\text{t}), \mathbf{X}_{it}]  
            \quad \because \text{ A1, C3} \nonumber \\
\mathbb{E}[Y_{it}(0) \mid H_{it}, g(\text{t}), \mathbf{X}_{it}] &= \mathbb{E}[Y_{it}(0) \mid Z_{it}=0, H_{it}, g(\text{t}), \mathbf{X}_{it}]  
            \quad \because \text{ C2} \nonumber \\
            &= \mathbb{E}[Y_{it} \mid Z_{it}=0, H_{it}, g(\text{t}), \mathbf{X}_{it}]  
            \quad \because \text{ A1, C3} \nonumber
\end{align}
therefore:
\begin{align}
\text{RR}^{\text{DLNM}} &= \frac{\mathbb{E}[Y_{it}(1) \mid H_{it}, g(\text{t}), \mathbf{X}_{it}]}
                    {\mathbb{E}[Y_{it}(0) \mid H_{it}, g(\text{t}), \mathbf{X}_{it}]} \nonumber \\
            &= \frac{\mathbb{E}[Y_{it} \mid Z_{it}=1, H_{it}, g(\text{t}), \mathbf{X}_{it}]}
                    {\mathbb{E}[Y_{it} \mid Z_{it}=0, H_{it}, g(\text{t}), \mathbf{X}_{it}]}  
            \quad \because \text{ Identification above} \nonumber \\
            &= \exp(\gamma) 
            \quad \because \text{Correct model specification.}
\end{align}
This identifies the ``added effect'' of heatwaves beyond continuous temperature effects captured by $h(H_{it}, \text{lag})$. Correct model specification requires the correct modeling of both $h(\cdot)$ and $g(\cdot)$.

\subsubsection*{\textbf{Matching Design}}
Recall the model:
\begin{equation}
\log(\mathbb{E}[Y_{it}]) = \alpha_i + \beta Z_{it}
\end{equation}
Recall the assumptions:
\begin{enumerate}
\item[D2. ] Within-unit temporal exchangeability: $\mathbb{E}[Y_{it}(0) \mid X_i=x_i] = \mathbb{E}[Y_{it'}(0) \mid X_i=x_i]$ for matched control periods $t'$;
\item[D3. ] Treatment overlap: Both treated and control periods exist for unit $i$;
\item[D4. ] No carry-over effects: $Y_{it}(z_{it})\perp\!\!\!\perp Z_{i,s}$ for $s \neq t$;
\item[D5. ] Stable unit composition: Same population at risk across matched periods.
\end{enumerate}
Then, one can identify from the data:
\begin{align}
\mathbb{E}[Y_{it}(1) \mid X_i=x_i] &= \mathbb{E}[Y_{it}(1) \mid Z_{it}=1, X_i=x_i]  
            \quad \because \text{ Treatment assignment deterministic at } t \nonumber \\
            &= \mathbb{E}[Y_{it} \mid Z_{it}=1, X_i=x_i]  
            \quad \because \text{ A1, D3} \nonumber \\
\mathbb{E}[Y_{it}(0) \mid X_i=x_i] &= \mathbb{E}[Y_{it'}(0) \mid X_i=x_i]  
            \quad \because \text{ D2, D4, D5} \nonumber \\
            &= \mathbb{E}[Y_{it'}(0) \mid Z_{it'}=0, X_i=x_i]  
            \quad \because \text{ Treatment assignment deterministic at } t' \nonumber \\
            &= \mathbb{E}[Y_{it'} \mid Z_{it'}=0, X_i=x_i]  
            \quad \because \text{ A1, D3} \nonumber
\end{align}
Therefore:
\begin{align}
\text{RR}^{\text{CC}}_i &= \frac{\mathbb{E}[Y_{it}(1) \mid X_i=x_i]}
                    {\mathbb{E}[Y_{it}(0) \mid X_i=x_i]} \nonumber \\
            &= \frac{\mathbb{E}[Y_{it} \mid Z_{it}=1, X_i=x_i]}
                    {\mathbb{E}[Y_{it'} \mid Z_{it'}=0, X_i=x_i]}  
            \quad \because \text{ Identification above } \nonumber \\
            &= \exp(\beta) 
            \quad \because \text{Correct model specification}
\end{align}
Within-unit comparison eliminates time-invariant confounding but requires temporal exchangeability.

\subsection*{\textbf{Identification of the Synthetic Control Estimator}}

For the sake of completeness here we report the identification of the causal effect, firstly proposed by \cite{abadie2010synthetic}. 
The identification strategy presented below has been reformulated to accommodate the SUTNVA assumption and align with the novel potential outcomes framework introduced in Section \ref{sec:method}.
Fix a treated unit $i\in\mathcal{Z}_t$ that first experiences a heatwave after period $T_0$. Let untreated potential outcomes satisfy (Equation~\ref{eq:dgp})
\begin{equation}
\label{eq:sc_dgp}
Y_{it}(0,0)=\delta_t+\mu_i+f_t^\top \lambda_t+\varepsilon_{it},
\qquad \mathbb{E}[\varepsilon_{it}]=0,
\end{equation}
where $\delta_t$ are common time shocks, $\mu_i$ a unit effect, $f_t\in\mathbb{R}^r$ time factors, $\lambda_t\in\mathbb{R}^r$ unit loadings, and $\varepsilon_{it}$ idiosyncratic errors. For $i\in\mathcal{Z}_t$, the synthetic control (SC) uses donor weights $w^{\,i}=\big(\,w^{\,i}_j\,\big)_{j\in\mathcal{N}_0^t}\in\Delta^{|\mathcal{N}_0^t|-1}$ chosen from the pre-period to form, for $t>T_0$, the post-treatment counterfactual
\begin{equation}
\label{eq:sc_cf}
\widehat{Y}_{it}(0,0)=\sum_{j\in\mathcal{N}_0^t}\widehat w^{\,i}_j\,Y_{jt}.
\end{equation}

\paragraph{\textbf{Proposition (Identification)}.}
Under SUTVA and assumptions \ref{ass:overlap}--\ref{ass:shocks}, we have that
\begin{equation}
\mathbb{E}\!\left[\widehat{Y}_{it}(0,0)\right]=Y_{it}(0,0)
\;\;\Longrightarrow\;\;
\mathbb{E}\!\left[\widehat{\tau}_{it}\right]=\tau_{it}\quad (t>T_0),
\end{equation}
where $\widehat{\tau}_{it}=Y_{it}(1, \cdot)-\widehat{Y}_{it}(0,0)$.% and $\tau_{it}=Y_{it}(1)-Y_{it}(0)$.

\paragraph{Proof (sketch).}
Following \cite{abadie2010synthetic}, consider the population synthetic path formed with $w^{\ast,i}$. Using the untreated DGP \eqref{eq:sc_dgp} and Assumption~A3
\begin{align*}
\sum_{j\in\mathcal{N}_0^t} \widehat w^{i}_j\,Y_{jt}(0,0)
&= \sum_{j\in\mathcal{N}_0^t} \widehat w^{i}_j\big(\delta_t+\mu_j+\lambda_t^\top f_j+\varepsilon_{jt}\big) \\
&= \delta_t + \sum_{j} \widehat w^{i}_j\mu_j + \lambda_t^\top \sum_{j} \widehat w^{i}_j f_j + \sum_{j} \widehat w^{i}_j \varepsilon_{jt}.
\end{align*}
By Assumption~A1, $\sum_j \widehat w^{i}_j\mu_j=\mu_i$ and $\sum_j \widehat w^{i}_j f_j=f_i$, hence
\begin{equation*}
\sum_{j\in\mathcal{N}_0^t} \widehat w^{i}_j\,Y_{jt}(0,0)
= \delta_t+\mu_i+\lambda_t^\top f_i+\sum_j \widehat w^{i}_j\varepsilon_{jt}
= Y_{it}(0,0)-\varepsilon_{it}+\sum_j \widehat w^{i}_j\varepsilon_{jt}.
\end{equation*}
Taking expectations and using Assumption~A4 gives
\begin{equation*}
\mathbb{E}\!\left[\sum_{j\in\mathcal{N}_0^t} \widehat w^{i}_j\,Y_{jt}(0,0)\right]
= Y_{it}(0,0).
\end{equation*}
If the estimated weights $\widehat w^{\,i}$ achieve negligible pre-period discrepancy, then $\mathbb{E}[\widehat{Y}_{it}(0,0)]\approx Y_{it}(0,0)$. For $t>T_0$,
\begin{align*}
\mathbb{E}[RRT^{SC}]
&=\mathbb{E}\!\left[\frac{Y_{it}(1, \cdot)}{\widehat{Y}_{it}(0,0)}\right]
= \mathbb{E}\!\left[ RRA^{SC} \times \frac{Y_{it}(0,0)}{\widehat{Y}_{it}(0,0)}\right] \\
&= RRT^{SC} \times \mathbb{E}\!\left[\frac{Y_{it}(0,0)}{\widehat{Y}_{it}(0,0)}\right]
\approx RRT^{SC}.
\end{align*}
\hfill$\square$

% \paragraph{Bias decomposition under approximate overlap.}
% Define
% \begin{equation}
% \Delta_\mu=\mu_i-\sum_{j\in\mathcal{N}_0^t}\widehat w^{\,i}_j\,\mu_j,
% \qquad
% \Delta_f=f_i-\sum_{j\in\mathcal{N}_0^t}\widehat w^{\,i}_j\,f_j.
% \end{equation}
% Then, using the untreated DGP \eqref{eq:sc_dgp} and Assumption~A3,
% \begin{align*}
% \mathbb{E}\!\big[\widehat{Y}^{\,i}_t(0)-Y_{it}(0)\big]
% &= \mathbb{E}\!\left[\sum_{j}\widehat w^{\,i}_j Y_{jt}(0) - \big(\delta_t+\mu_i+\lambda_t^\top f_i+\varepsilon_{it}\big)\right] \\
% &= \left(\sum_{j}\widehat w^{\,i}_j\mu_j - \mu_i\right)
% \;+\;
% \lambda_t^\top \left(\sum_{j}\widehat w^{\,i}_j f_j - f_i\right)
% \;+\;
% \mathbb{E}\!\left[\sum_{j}\widehat w^{\,i}_j \varepsilon_{jt} - \varepsilon_{it}\right] \\
% &= \Delta_\mu + \lambda_t^\top \Delta_f
% \;+\;
% \mathbb{E}\!\left[\sum_{j}\widehat w^{\,i}_j \varepsilon_{jt} - \varepsilon_{it}\right].
% \end{align*}
% Hence, small pre-period imbalance in $(\mu_i,f_i)$ together with the mean-zero shock condition implies a small systematic bias post-treatment.

\subsection*{\textbf{Identification of the Spatially--Augmented SC under a Spatial DGP}}

Fix a treated unit $i\in\mathcal{Z}_t$ whose first exposure occurs after $T_0$.
Untreated potential outcomes follow the spatial interactive fixed--effects DGP
\begin{equation}
\label{eq:spatial_dgp}
Y_{it}(0,0)=\alpha_i+\delta_t+ f_t^\top B_i+u_{it}+\varepsilon_{it},
\end{equation}
where (i) $\bm\alpha=(I-\varpi_\alpha W)^{-1}\bm\nu_\alpha$ with row--stochastic $W$ and $|\varpi_\alpha|<1$; 
(ii) $(\delta_t)$ is covariance--stationary; 
(iii) $f_t\in\mathbb{R}^{r}$ is low--dimensional (e.g., AR(1)) and $B_i$ are spatially smoothed loadings (e.g., $(I-\varpi_B W)^{-1}\bm\nu_B$ with $|\varpi_B|<1$); 
(iv) $u_{it}=\rho_u (Wu)_{i,t-1}+\epsilon_{it}$ with $|\rho_u|<1$, $\mathbb{E}[u_{it}]=0$, and summable auto--covariances; 
(v) $\varepsilon_{it}$ is mean--zero measurement noise with bounded variance. 
Let $s_i(t)\in\{0,1\}$ indicate spillover exposure; assume $s_i(t)=0$ for $t\le T_0$ (no anticipation).
For $t>T_0$, we exclude from the donor pool any unit with $z_j(t)=1$ or $s_j(t)=1$.

For treated unit $i$, SA--SC forms weights $w^{\,i}\in\Delta^{\mathcal{N_0^{\ast t}}}$ (over the pre--treatment donor set) via a logistic--normal map
\[
w^{\,i}=\mathrm{softmax}(\eta-\bar\eta),\qquad 
\eta=-\varsigma D_{(ij)}+\tau_\eta z,\quad z\sim N(0,I_K),
\]
with distances $D_{i\cdot}\ge 0$, intensity $\varsigma\ge 0$, and scale $\tau_\eta>0$. 
The post--treatment counterfactual for $t>T_0$ is
\begin{equation}
\label{eq:cf_sasc}
\widehat Y_{it}(0,0)=\sum_{j\in\mathcal N^{*t}_{0}}\widehat w^{\,i}_j\,Y_{jt}.
\end{equation}

\paragraph{Mapping to the classical SC structure.}
Write \eqref{eq:spatial_dgp} as a standard interactive fixed--effects model by setting
\[
\mu_i:=\alpha_i,\qquad 
\lambda_t:=f_t,\qquad 
\text{and}\quad \varepsilon^{\dagger}_{it}:=u_{it}+\varepsilon_{it}.
\]
Then
\begin{equation}
\label{eq:ife_rewrite}
Y_{it}(0,0)=\delta_t+\mu_i+\lambda_t^\top B_i+\varepsilon^{\dagger}_{it},
\end{equation}
which is the Abadie--Diamond--Hainmueller untreated DGP with spatially structured $(\mu_i,B_i)$ and a weakly dependent error $\varepsilon^{\dagger}_{it}$. 
Hence the spatial DGP is a special case of the SC framework: spatial smoothing affects feasibility (support) but not the identification logic.

% \paragraph{Assumptions.}
% (A1) \{SUTVA and no anticipation:} for $t\le T_0$, $Y_{jt}=Y_{jt}(0,0)$ for all $j$. \\
% (A2) \{Buffered donors post--treatment:} for $t>T_0$, $\mathcal N^{t}_{0,\mathrm{buf}}\subseteq\{j: z_j(t)=0,\ s_j(t)=0\}$. \\
% (A3) \emph{Convex feasibility (support):} there exist $w^{\ast,i}\in\Delta^{K-1}$ with 
% $\mu_i=\sum_j \widehat w^{i}_j\mu_j$ and $B_i=\sum_j \widehat w^{i}_j B_j$. \\
% (A4) \emph{Mean--zero weak dependence:} $\mathbb{E}[\varepsilon^{\dagger}_{it}]=0$, 
% $\sup_{it}\mathbb{E}[(\varepsilon^{\dagger}_{it})^2]<\infty$, and auto--covariances of $\varepsilon^{\dagger}_{it}$ are summable (implied by $|\rho_u|<1$ and bounded innovations). \\
% (A5) \emph{Bounded distance penalty:} $\max_j \lambda\, d^{(i)}_j<\infty$ (the anchor does not diverge with $T_0$).

\paragraph{\textbf{Proposition (Identification).}}
Under SUTNVA and \ref{ass:overlap}--\ref{ass:shocks},
\[
\mathbb{E}\!\left[\widehat Y_{it}(0,0)\right]=Y_{it}(0,0)
\;\;\Longrightarrow\;\;
\mathbb{E}\!\left[\widehat RRT ^{SA-SC}\right]=RRT ^{SA-SC}\quad (t>T_0),
\]
where $\widehat{RRT}^{SA-SC}=Y_{it}-\widehat Y_{it}(0,0)$ and $RRT ^{SA-SC}=Y_{it}-Y_{it}(0,0)$.

\paragraph{Proof (sketch).}
Using \eqref{eq:ife_rewrite} and (A3),
\begin{align*}
\sum_{j\in\mathcal N^{\ast t}_0} \widehat w^{i}_j Y_{jt}(0,0)
&=\sum_j \widehat w^{i}_j\big(\delta_t+\mu_j+\lambda_t^\top B_j+\varepsilon^{\dagger}_{jt}\big) \\
&=\delta_t+\sum_j \widehat w^{i}_j\mu_j+\lambda_t^\top\sum_j \widehat w^{i}_j B_j+\sum_j \widehat w^{i}_j \varepsilon^{\dagger}_{jt} \\
&=\delta_t+\mu_i+\lambda_t^\top B_i+\sum_j \widehat w^{i}_j \varepsilon^{\dagger}_{jt} \\
&=Y_{it}(0,0)-\varepsilon^{\dagger}_{it}+\sum_j \widehat w^{i}_j \varepsilon^{\dagger}_{jt}.
\end{align*}
Taking expectations and invoking (A4) yields 
$\mathbb{E}\big[\sum_j \widehat w^{i}_j Y_{jt}(0,0)\big]=Y_{it}(0,0)$.
If the estimated weights $\widehat w^{\,i}$ achieve negligible pre--fit discrepancy, the bounded distance anchor (A5) implies $\widehat w^{\,i}\approx w^{i}$ as $T_0$ grows, hence $\mathbb{E}[\widehat Y_{it}
(0,0)]\approx Y_{it}(0,0)$.
For $t>T_0$, by (A2) donors remain unexposed (including to spillovers), so
\begin{align*}
\mathbb{E}[RRT^{SA-SC}]
&=\mathbb{E}\!\left[\frac{Y_{it}(1, \cdot)}{\widehat{Y}_{it}(0,0)}\right]
= \mathbb{E}\!\left[ RRA^{SA-SC} \times \frac{Y_{it}(0,0)}{\widehat{Y}_{it}(0,0)}\right] \\
&= RRT^{SA-SC} \times \mathbb{E}\!\left[\frac{Y_{it}(0,0)}{\widehat{Y}_{it}(0,0)}\right]
\approx RRT^{SA-SC}.
\end{align*}

\section{Implementation}
\label{sec:implementation}
%\leo{I'm using 60 days for the application. Should we write different implementation sections for the simulation and application?}
For every identified heatwave episode we constructed a donor pool made of counties that, on the corresponding days, (i) were not in a heatwave, (ii) had complete outcome rates for the 60 days before the start of the episode, and (iii) were not in a radius of 20km from the treated county.
We use a 20km buffer based on typical emergency medical service catchment areas, though sensitivity to this choice warrants future investigation
The 60 day pre-treatment period was used to learn the temporal pattern of mortality (in the simulations) or hospitalization rates (in the application) in the treated county and to match it to a weighted combination of donor counties. 
We then estimated the spatially augmented Bayesian synthetic control model described in Section \ref{sec:method}. 
In this model the counterfactual rate for the treated county is expressed as a convex combination of rates from donor counties, and the donor weights receive a spatial prior that favors nearby counties. 
Estimation was carried out in a Bayesian framework using Hamiltonian Monte Carlo with four chains and 5,000 iterations per chain. For every treated county and every treated day, we drew from the posterior predictive distribution of the counterfactual rate that would have been observed in the absence of the heatwave. 
The causal effect was summarized as a posterior mean relative risk of rate and a 95 percent credible interval.

\section{Simulation Details and Extensions}

Consider a balanced panel with units $i=1,\ldots,N$ indexing U.S. counties and times $t=\{1,\ldots,T\}$ indexing calendar days. Let $\mathcal T_-=\{1,\ldots,T_0\}$ denote the pre-treatment window ($|\mathcal{T_-}| = 20 $) and $\mathcal T_+=\{T_0+1,\ldots,T\}$ the post-treatment window ($|\mathcal{T_+}| = 10 $). 
In the reported simulations, $T_0$ is set to {1 June} of the year $y$ sampled from 2000--2016 within a New England FIPS subset.
Spatial dependence is encoded by an $N\times N$ sparse, row-stochastic matrix $W$. 
Two constructions are admissible: 
\begin{itemize}
    \item $\bm W_{\text{geo}}$ a $k$-nearest-neighbour (k-NN) kernel on county centroids (lon/lat), with $w_{ij}\propto \exp(-D_{ij}/h)$ on the $k$ nearest neighbours, symmetrised and row-standardised;
    \item $\bm W_{\text{adj}}$ a polygon contiguity matrix, with 1 if two counties shares a border and zero otherwise.
\end{itemize}

In the outcome's data generating process (DGP), we employ a k-NN kernel with $k=4$ and bandwidth $h=0.5h_0$, where $h_0$ is the median inter-point distance computed from 4-NN; spillovers are computed with queen contiguity using the identity-islands convention.
The potential log-rates in absence of treatment and spillover contamination will follow
\begin{equation}\label{eq:dgp-y0}
Y_{it}(0, 0)=\alpha_i+\delta_t+ f_tB_{i}+u_{it}+\varepsilon_{it}\;,
\end{equation}
Where $\alpha_i$ represents a spatially smoothed unit-level fixed effect, $\delta_t$ represents a time level fixed effect, $f_t$ is an AR(1) 2-dimensional temporal latent factor which represents the evolution of the outcome through times, and $B_i$ is the $i$-th row of a $N\times 2$ matrix which represents a spatially smoothed factor loading. 
Finally, $u_{it}$ is the part of idiosyncratic error that propagates spatially, and $\varepsilon_{it}$ is the iid error. 
Unit effects and loadings are spatially smoothed via a SAR-type operator:
\begin{equation}\label{eq:sar-smooth}
\alpha=(I-\varpi_\alpha \bm W_{\text{geo}})^{-1}\nu_\alpha,\qquad \nu_\alpha\sim \text{N}(0,\sigma_\alpha^2 I),
\end{equation}
where $\alpha$ is the $N$-dimensional vector whose $i$-th unit is $\alpha_i$, and, $\varpi_\alpha$ . 
Then, for each factor $g=1,2$, we define the $N$-dimentional spatial loading $B_g$ as follows:
\begin{equation}
B_{g}=(I-\varpi_b \bm W_{\text{geo}})^{-1}\nu_{b,g},\qquad \nu_{b,g}\sim \text{N}(0,\sigma_b^2 I).
\end{equation}
 
Temporal dynamics comprise $G$ latent factors and a common component:
\begin{equation}
\begin{split}
f_{t,g}=\phi_{f,g}f_{t-1,g}+\epsilon_{t,g},\quad \epsilon_{t,g}\sim\text{N}(0,\sigma_{f,g}^2), \quad \forall g = 1, \dots, G\\
\gamma_t=\phi_\gamma\gamma_{t-1}+\xi_t,\quad \xi_t\sim\text{N}(0,\sigma_\gamma^2), \quad \forall t = 1, \dots, T.
\end{split}
\end{equation}
Idiosyncratic deviations propagate spatially as
\begin{equation}
u_{it}=\rho_uW_{\text{geo}}u_{it-1}+\zeta_t,\qquad \zeta_t\sim\text{N}(0,\sigma_u^2 I),
\end{equation}
and measurement noise is i.i.d.\ $\varepsilon_{it}\sim\mathcal N(0,\sigma_\varepsilon^2)$.
Daily heat $h_{it}$ is scaled as z-pre using mean and standard deviation computed over $\mathcal T_-$, yielding $\tilde h_{it}$. The treatment effect is active only for treated units in the post period and increases non-linearly with heat:
\begin{equation}\label{eq:treatment}
\tau_{it} = \tau_0{\exp(\kappa \tilde h_{it})-1}  \qquad {i \in \mathcal{N}_1}, {t\ge T_0}
\end{equation}

%\paragraph{Spillovers onto controls (post only).}
As the exposure to heatwave possibly could generate spillover effect on units in the proximity of the treated one, we generate what part of the treatment effect is spilling on untreated units. 
Thus we will define the spillover effect received by untreated unit $j$ from treated unit $i$ located at degree $s_{ij} = s$ of separation as 
\begin{equation}
    \psi^{s}_{jt} = \chi^s W_{\text{adj}}^{s} \tau_{it}.
\end{equation}
Where $W_{\text{adj}}^s$ represent the set of units located at step $s_{ij}$ from unit i and $\chi^s$ represent the amount of treatement effect that spills out from treated unit. 
In our example, we focus to spillover for units located at 1-degree and 2-degree of separation, with $\chi = (0.7, 0.4)$. 
We consider multiplicative effect from the treatment onset, our observed log-rates for mortality are 

\begin{equation}
    \begin{cases}
        Y_{it}=Y_{it}(0, 0)\times\tau_{it}, & Z_{it} = (1,\cdot) \\ 
        Y_{it}=Y_{it}(0, 0)\times\psi^s_{it}, & Z_{it} = (0,1)\\
        Y_{it}=Y_{it}(0, 0), & Z_{it} = (0,0)
    \end{cases}
\end{equation}

% Spillovers are added to both $Y_{it}(0)$ and $y_{it}$ for controls only, using the numeric weights in $W$ (not merely binary adjacency). This allows first- and second-order diffusion while preserving the treated--control partition.

% so that observed log-rates are
% \begin{equation}\label{eq:observed}
% Y_{it}=Y_{it}(0, 0)+\tau_{it}.
% \end{equation}
Background spatial dependence arises from SAR-smoothed unit effects and loadings combined with dynamic propagation in $u_t=\rho_u W u_{t-1}+\zeta_t$, generating spatial clustering and persistence that mirror demographic sorting, commuting sheds, and jurisdictional proximity. Temporal persistence and common shocks are represented by autoregressive latent factors and a shared AR(1) component, accommodating seasonal structure and meteorological anomalies while allowing heterogeneous factor loadings across space. The exposure--response is non-linear in heat intensity: the exponential term in $\tilde h_{it}$ yields limited effects for mild deviations and acceleration at extreme quantiles, consistent with threshold-like responses. Finally, the framework explicitly contaminates control trajectories via geographic spillovers generated by treated units, a mechanism that reflects diffusion phenomena, behavioural displacement, service congestion, or networked interactions, prevalent in metropolitan systems and directly challenges synthetic-control identification in realistic settings.

%\subsection{Optional calibration to external baselines.}
To align pre-treatment levels with external crude rates (per-100k-year converted to daily), compute a county-specific shift
\begin{equation}
r_i=\frac{1}{|\mathcal{T_-|}}\sum_{t\in\mathcal T_-}\big[\log r^{\text{target}}_{it}-Y_{it}(0,0)\big],
\end{equation}
and apply
\begin{equation}
Y_{it}^\star(0,0)=Y_{it}(0,0)+r_i.
\end{equation}
leaving $\tau_{it}$ unchanged. Calibration restores realistic scale and cross-county heterogeneity in levels without altering treatment structure.

\paragraph{Parameterisation used in the reported runs.}
Unless stated otherwise: $K=2$; $\lambda_\alpha=0.5$, $\lambda_b=0.5$; $\rho_u=0.2$; $\sigma_\varepsilon=0.05$; $\sigma_u=0.15$; $\phi_\gamma=0.68$, $\sigma_\gamma=0.11$; $\phi_{f,k}=0.85
$, $\sigma_{f,k}=0.08$ for $k=1,2$; $\tau_0=5.7$, $\kappa=0.65$. Heat scaling is z-pre on the heat index variable. The outcome DGP uses k-NN $\bm W_{geo}$ with $k=4$ and $h=0.5h_0$; spillovers use queen contiguity with identity islands.

% \paragraph{Code-to-notation crosswalk.}
% \begin{center}
% \begin{tabular}{ll}
% \hline
% \textbf{R object / argument} & \textbf{Notation / role} \
% \hline
% \texttt{panel_dt[county,date]} & indices $(i,t)$, panel ordering \
% \texttt{W} from \texttt{build_W_kernel}/\texttt{build_W_contiguity} & spatial weights $W$ \
% \texttt{lambda_alpha}, \texttt{lambda_b} & $\lambda_\alpha$, $\lambda_b$ (SAR smoothing) \
% \texttt{rho_u}, \texttt{sigma_u}, \texttt{sigma_eps} & $\rho_u$, $\sigma_u$, $\sigma_\varepsilon$ \
% \texttt{phi_f}, \texttt{sigma_f}, \texttt{phi_gamma}, \texttt{sigma_gamma} & $\phi_{f,k}$, $\sigma_{f,k}$, $\phi_\gamma$, $\sigma_\gamma$ \
% \texttt{scale_heat}/\texttt{heat_var}, \texttt{tau0}, \texttt{kappa} & $\tilde h_{it}$, $\tau_0$, $\kappa$ \
% \texttt{treated_counties}, \texttt{T0} & treatment assignment, $\text{post}t$ \
% \texttt{add_spillover_in_controls(..., alpha1, alpha2, W_{geo}pill)} & $\alpha_1$, $\alpha_2$, $W^2$ \
% \texttt{calibrate_sim_to_baseline} & $s_i$, $Y{it}^\star(0)$, $y_{it}^\star$ \
% \hline
% \end{tabular}
% \end{center}

In each replication, the year $y$ is sampled from 2000--2016; $T_0=$1 June; the random seed is $61116+jj$, where $jj$ indexes the replicate. For transparency, report $N=217$, $T=30$, $|\mathcal T_-|=20$, date span, seed, and whether calibration and spillovers are active for each run.

\subsection*{\textbf{Simulation Metrics}}

We assess each estimator by the bias and Mean Squared Error (MSE) of the
imputed {control} potential outcomes for treated units, averaged over the
post–treatment horizon and across treated units. Let L be the total number of replication and $T_1$ and $T_2$ be the extremes of the time span in which we are evaluating the estimator (e.g.: if we evaluate the pre-treatment $T_1=0 \text{ and } T_2 = T_0$. Formally,
\[
\mathrm{Bias}
= \frac{1}{T_2 - T_1}\sum_{t=T_1}^{T_2} \mathrm{Bias}_{it}
= \frac{1}{T_2 - T_1}\sum_{t=T_1}^{T_2}\frac{1}{L}
\big(\big|\widehat Y_{it}(0,0) - Y_{it}(0,0)\big|\big),
\]
and
\[
\mathrm{RMSE}
= \sqrt{\frac{1}{T_2 - T_1}\sum_{t=T_1}^{T_2} \mathrm{MSE}_{it}}
= \sqrt{\frac{1}{T_2 - T_1}\sum_{t=T_0+1}^{T}\sum_{i=1}^{N_1}
\big(\widehat Y_{it}(0,0) - Y_{it}(0,0)\big)^2},
\]
where $\widehat Y_{it}(0,0)$ denotes the imputed value and
$Y_{it}(0,0)$ the true (simulated) control potential outcome
for treated unit $i$ at time $t$. For Bayesian
procedures, bias and MSE are computed using posterior medians from the posterior
predictive distributions of the missing control potential outcomes.

We also report {average coverage probabilities} (ACP) of nominal
$95\%$ confidence/credible intervals. Let $C_{0.95}\!\big(\widehat Y_{it}(0,0)\big)$ be the corresponding $95\%$ interval; then
\[
\mathrm{ACP}
= \frac{1}{T_2 - T_1}\sum_{t=T_1}^{T_2}\mathrm{CP}_{it}
= \frac{1}{T_2 - T_1}\sum_{t=T_1}^{T_2}
\mathbb{I}\!\left\{\,Y_{it}(0,0) \in C_{0.95}\!\big(\widehat Y_{it}(0,0)\big)\right\}.
\]
We focus on these four methods because each yields a transparent measure of
uncertainty: classical confidence intervals for OLS and posterior credibility
intervals for the Bayesian approaches.

\subsection*{\textbf{Additional Simulation Results}}

\input{Tables/sim_sutva}

Table~\ref{tab:sim_res_sutva} reports the Monte Carlo results for counterfactual estimation when donor weights are estimated using the standard Synthetic Control (SC) and the Spatially--Augmented Synthetic Control (SA--SC), under the SUTVA assumption. 
When spillover effects are present but not explicitly accounted for, the absolute bias of SA--SC tends to increase more than that of SC. 
This pattern arises because the spatially augmented estimator assigns higher weight to geographically proximate control units, which are themselves partially affected by the treatment spillover, whereas the classical SC does not incorporate this spatial proximity. 
The same mechanism also influences the post--treatment RMSE, the coverage probability, and the average credible interval length, all of which deteriorate under unmodelled spillover exposure.
By contrast, when the data--generating process exhibits spatially correlated factor loadings and spatially dependent errors but no spillovers, SA--SC achieves lower bias and RMSE and higher coverage probability, indicating improved efficiency in capturing the underlying spatial structure. 
Under combined spatial confounding and spillover, the two estimators perform comparably in terms of bias and RMSE, although the coverage probability of SA--SC remains slightly inferior.
Overall, these results suggest that the spatial augmentation improves estimation in the presence of spatial confounding but may amplify bias when spillovers are ignored. 
Further research is warranted to delineate the practical conditions under which SA--SC provides systematic gains over standard SC in applied causal inference settings.

% \input{Tables/sim_sutnva}

% Table~\ref{tab:sim_sutnva} presents the results instead when we assume \ref{ass:sutnva} for both SC and SA--SC, following the proposal of \cite{grossi2025direct} for the selection of the donor pool. We can notice that once we restrict the donor pool to \textit{pure untreated units} the 

\begin{figure}
    \centering
    \includegraphics[width=\linewidth]{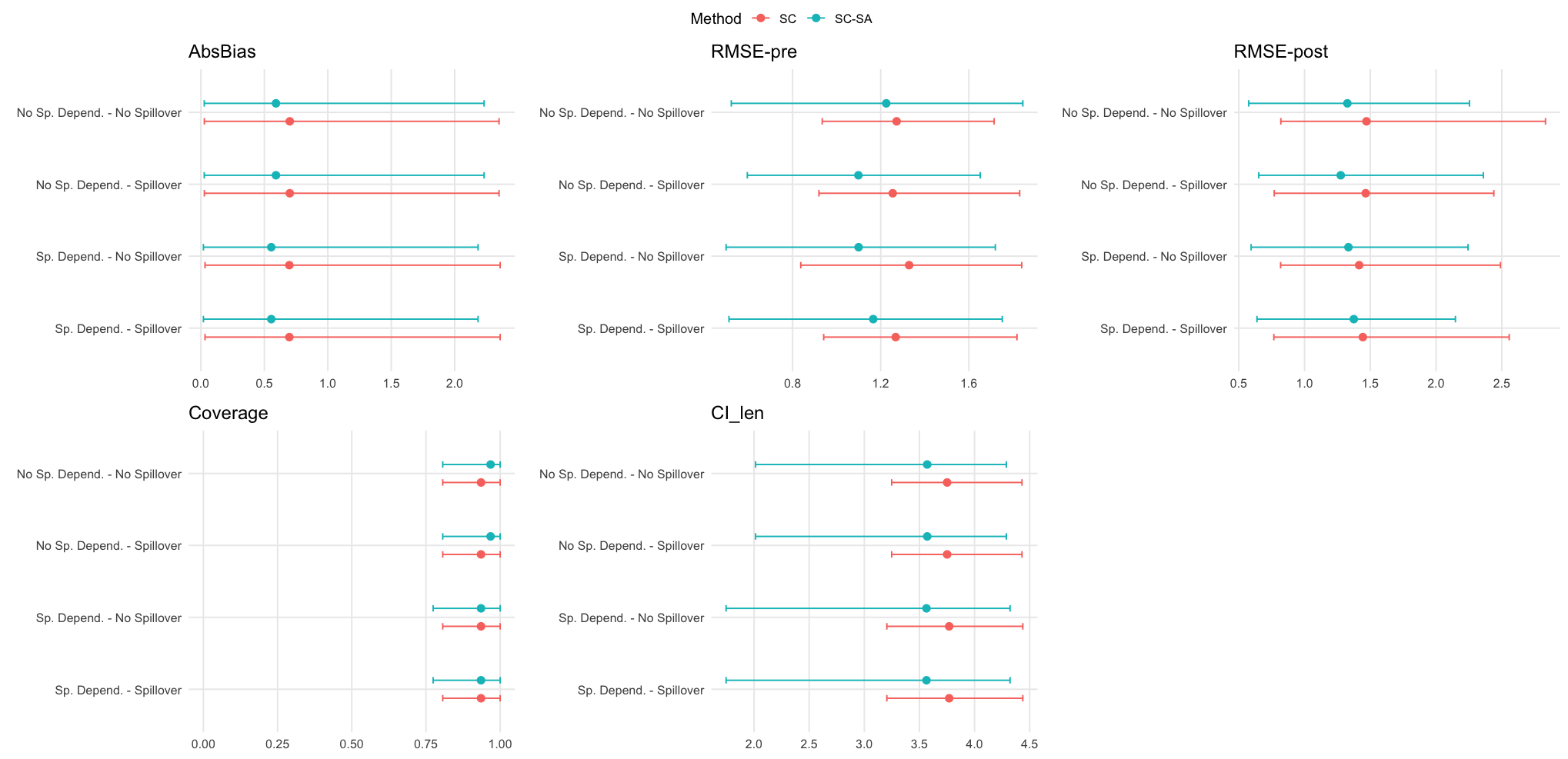}
    \caption{Monte Carlo performance under SUTNVA assumption across 100 replications for causal inference methods.  SC refers to \textit{synthetic control} while SA-SC refers to \textit{spatially augmented synthetic control}. The central dot represents the average value of the statistics, the upper and lower bound are the 2.5\% and 97.5\% quantiles of the statistics' distribution}
    \label{fig:
    sim_res_mis}
\end{figure}

\begin{figure}
    \centering
    \includegraphics[width=\linewidth]{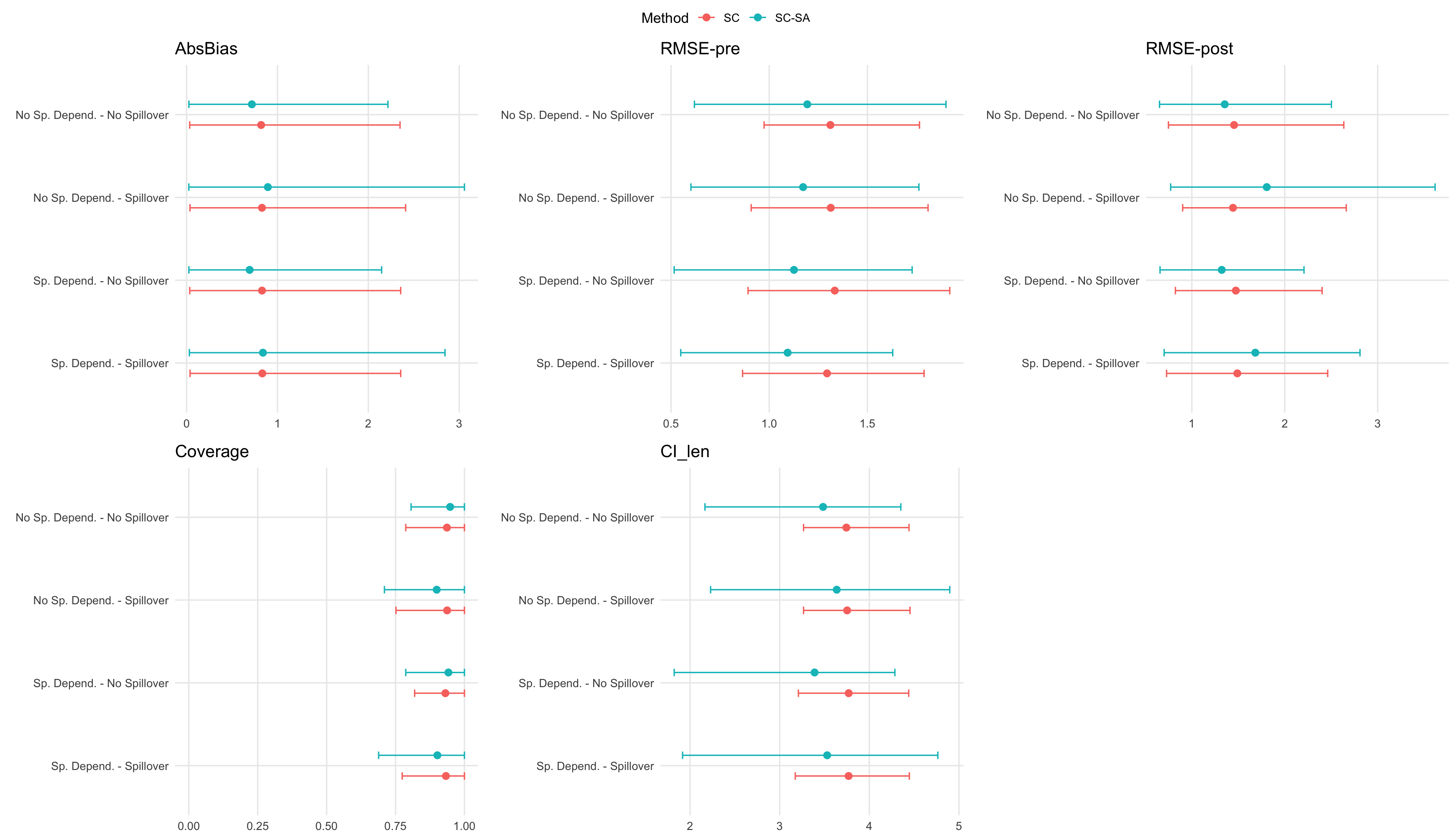}
    \caption{Monte Carlo performance under SUTVA assumption across 100 replications for causal inference methods.  SC refers to \textit{synthetic control} while SA-SC refers to \textit{spatially augmented synthetic control}.The central dot represents the average value of the statistics, the upper and lower bound are the 2.5\% and 97.5\% quantiles of the statistics' distribution}
    \label{fig:
    sim_res_mis}
\end{figure}

% In this table we report the results for counterfactual estimation when weights are estimated through SC and SA-SC. We can notice that in presence of spillover, when this is not identified, The absolute bias 
% for SA-SC increases more than SC, as it is privileging the control units that are closer to the treated one, while the classical SC is not considering this. This relation is affecting also the estimation of RMSE post-treatment and the estimation of coverage probability and the average CI length. Notably, an opposite effect can be seen when there are spatial structure in loadings and spatial errors: SC-SA provides improved estimates for the bias, RMSE, coverage and CI length. In presence of both spatial confounding and spillover, the estimates are comparable in our simulation in terms of bias, and RMSE, while the SA-SC coverage is inferior. Further research is needed on the topic to identify practical suggestions for the application fields of SA and SA-SC. 

\section{Additional Application Details and Results}

\subsection*{\textbf{Data Processing and Implementation}}

% For each identified heatwave episode, we constructed a donor pool composed of counties that were not experiencing a heatwave on the corresponding date, had complete outcome data for the sixty days prior to the heatwave onset, and were located outside the spatial neighborhood of the treated county to avoid spillover contamination. Among the eligible counties, we selected the twenty donors with covariate profiles most similar to the treated county, as measured by the Mahalanobis distance of pre-treatment covariates. We included the following covariates: percentage Black population, percent higher education, percent poverty, perecent owener occupied housing, medican home value, and median household income (see Table \ref{}). The pre-treatment period of sixty days was used to learn the temporal dynamics of hospitalization rates and to calibrate the donor weights that best reproduced the treated county’s pre-heatwave trajectory. All models were estimated within a Bayesian framework using Hamiltonian Monte Carlo (four chains, 1,000 iterations per chain, 500 warm-up).

Daily hospitalization rates were calculated by dividing the number of Medicare hospitalizations in the relevant ICD categories (\textit{heat-related illness} (ICD-9-CM 992.x; ICD-10-CM T67.x), \textit{fluid and electrolyte disorders} (276.x; E86.x), \textit{acute kidney failure} (584.x; N17.x), \textit{urinary tract infection} (599.0; N39.0), and \textit{septicemia and severe systemic infections} (038.x; A41.x)) by the number of Medicare beneficiaries residing in the county in that year. We focused on urban counties with populations exceeding 200,000, which ensured sufficient event counts and reduced the frequency of zero-hospitalization days. For counties with zero events, we applied a seven-day rolling average imputation to stabilize the rate series and avoid taking the logarithm of zero. The outcome variable used in all models was the natural logarithm of the imputed hospitalization rate. We also examined an alternative specification in Figure \ref{fig:plot-counties-rollavg}, where a five-day rolling average was applied to all days rather than imputing only zero-rate days.

In the application, both the standard SC and the SA-SC were implemented under the SUTNVA assumption. This assumption allows potential outcomes to depend on the treatment status of nearby counties and restricts donor pools to counties located outside the defined spillover radius. We adopted SUTNVA for both methods so that they would target the same estimand in the application, which ensures that their estimates are directly comparable.

The implementation procedure of the application followed the steps outlined in Appendix \ref{sec:implementation}. After identifying all eligible donor counties, we further selected the twenty donors whose pre-treatment covariate profiles most closely matched that of the treated county. Similarity was evaluated using the Mahalanobis distance computed from demographic and socioeconomic covariates measured during the pre-treatment period. The covariates included percentage Black population, percentage of adults with a higher education degree, percentage of individuals below the poverty line, percentage of owner-occupied housing, median home value, and median household income. Table \ref{tab:urban_rural_desc} summarizes the population and covariate characteristics.

\input{Tables/descriptive_table}

\subsection*{\textbf{Additional Application Results}}
\begin{figure}[h]
    \centering
    \includegraphics[width=0.9\linewidth]{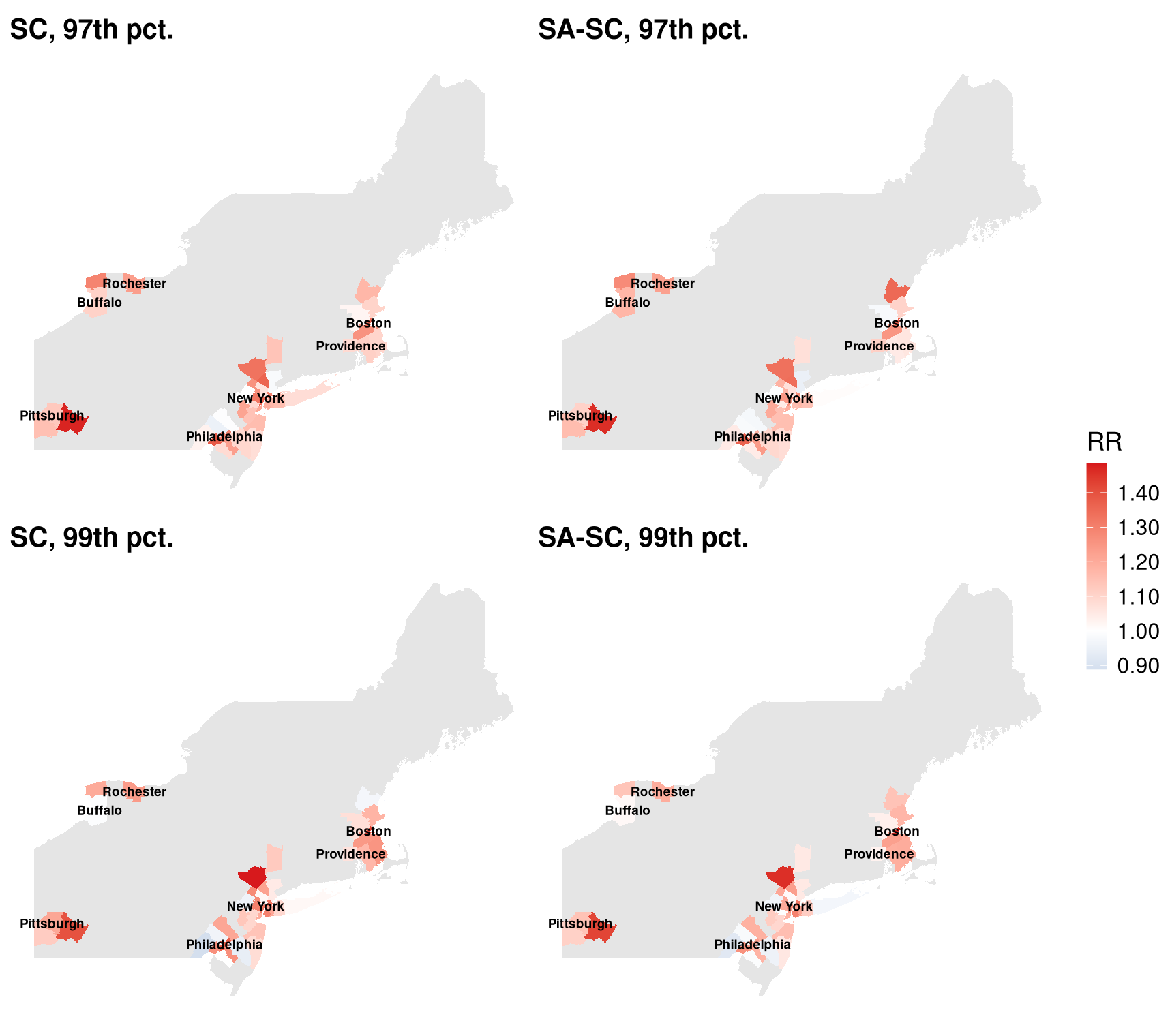}
    \caption{Posterior mean relative risk of heat-related hospitalizations during heatwaves versus non-heatwave days across urban Northeastern U.S. counties, 2000-2019. Heatwaves are defined as periods of at least two consecutive days exceeding the 97 or 99th percentile of  county-specific annual heat index values. Estimates correspond to first-season heatwaves with donor pools restricted to non-exposed counties outside  treated neighborhoods. Left panel: standard synthetic control (SC). Right panel: spatially-augmented synthetic control (SA-SC). Red indicates increased hospitalization risk (RR $>$ 1) during heatwaves; grey indicates counties excluded from estimation.}
    \label{fig:hw-definitions}
\end{figure}

\begin{figure}[h]
    \centering
    \includegraphics[width=0.85\linewidth]{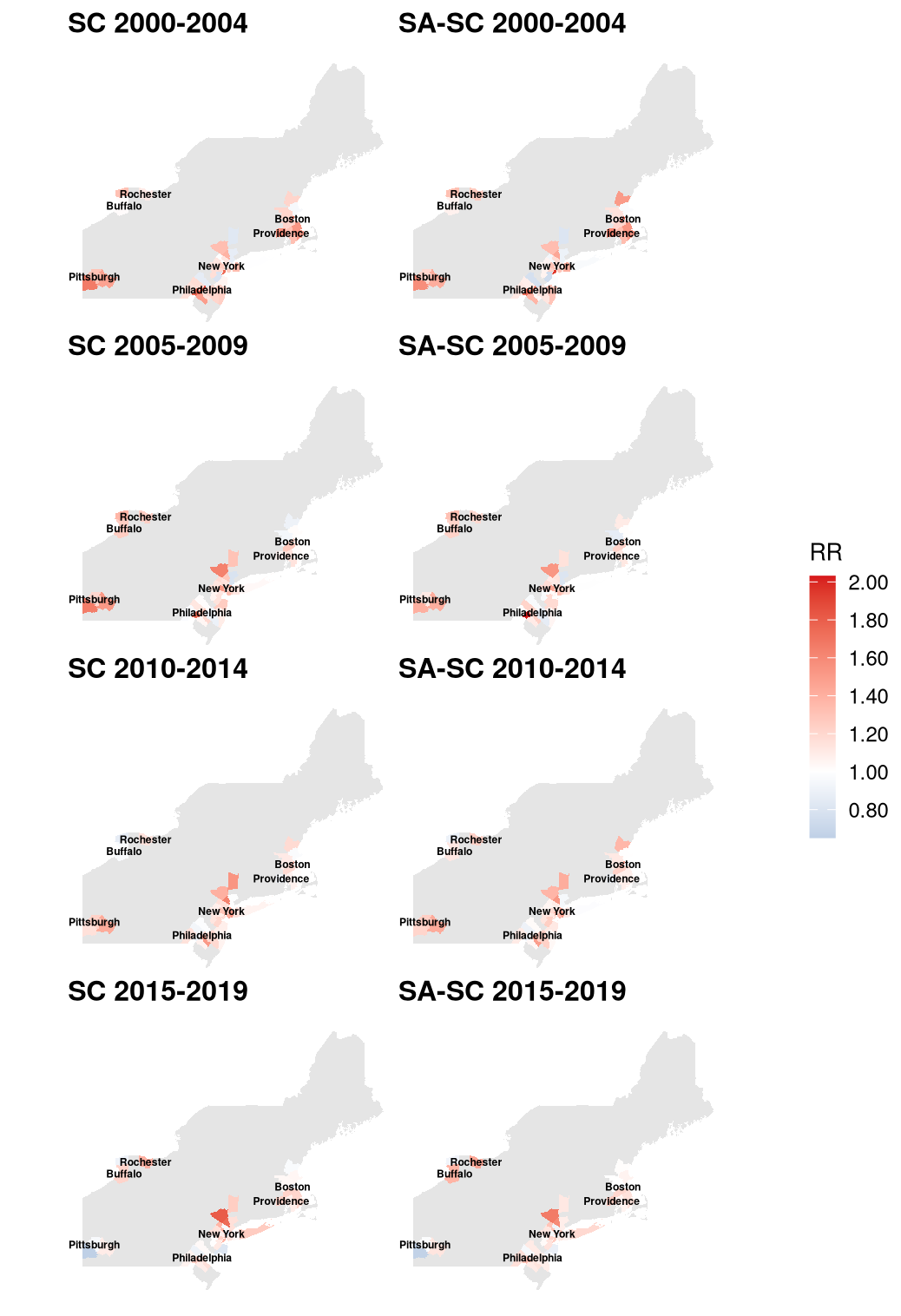}
    \caption{Posterior mean relative risk of heat-related hospitalizations during heatwaves versus non-heatwave days across urban Northeastern U.S. counties, across different five-year periods (2000-2004, 2005-2009, 2010-2014, 2015-2019). Heatwaves are defined as periods of at least two consecutive days exceeding the 95th percentile of  county-specific annual heat index values. Estimates correspond to first-season heatwaves with donor pools restricted to non-exposed counties outside treated neighborhoods. Counties that did not experience a heatwave or did not have eligible donor counties were not estimated. Left panel: standard synthetic control (SC). Right panel: spatially-augmented synthetic control (SA-SC). Red indicates elevated hospitalization risk (RR $>$ 1) during heatwaves; grey indicates counties excluded from estimation.}
    \label{fig:five-years-maps}
\end{figure}

\begin{figure}[h]
    \centering
    \includegraphics[width=\linewidth]{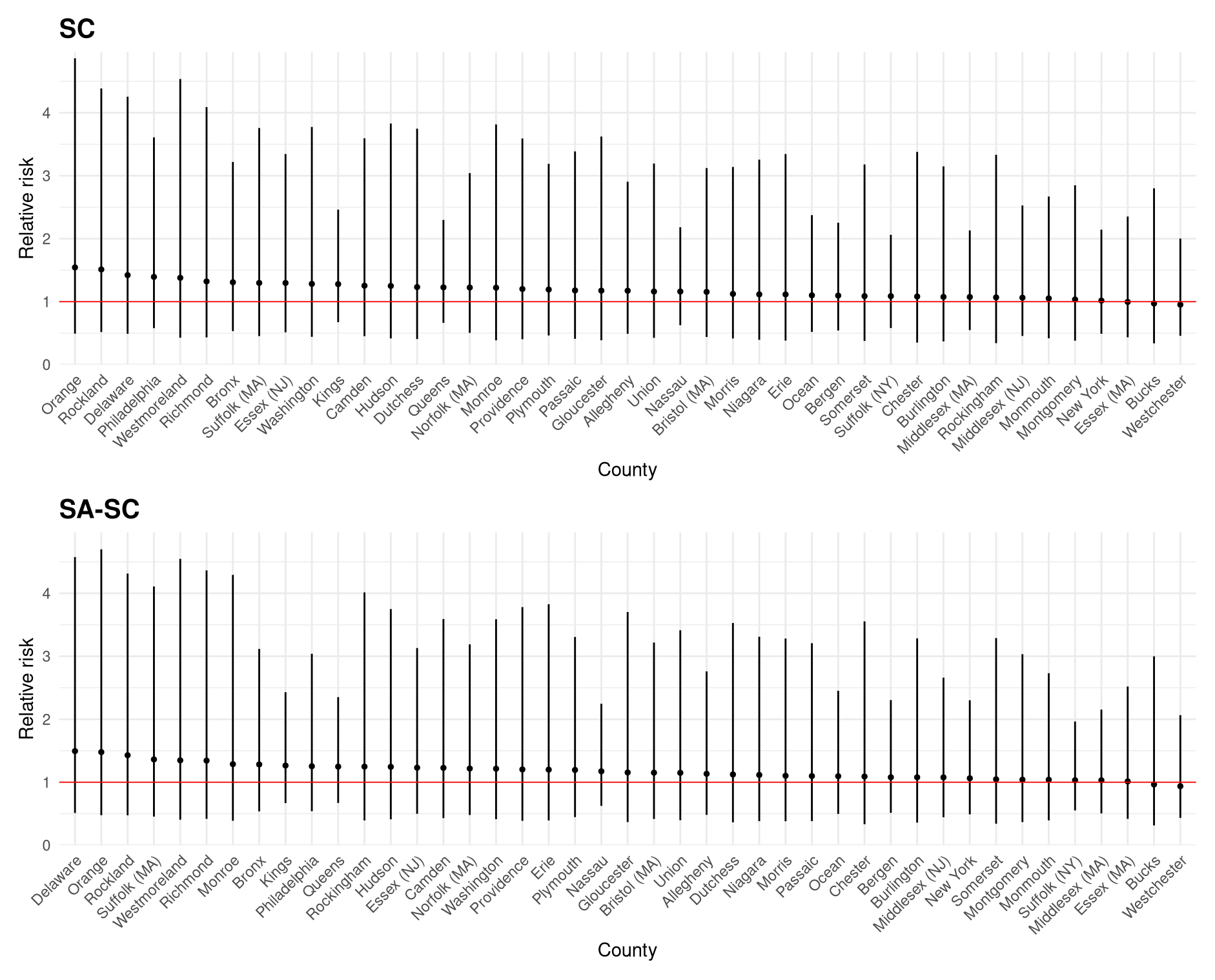}
    \caption{Posterior mean relative risk of heat-related hospitalizations during heatwaves versus non-heatwave days for each urban Northeastern U.S. county considered in our study, 2000-2019. Heatwaves are defined as periods of at least two consecutive days exceeding the 95th percentile of  county-specific annual heat index values. Estimates correspond to first-season heatwaves with donor pools restricted to non-exposed counties outside  treated neighborhoods. Top panel: standard synthetic control (SC). Bottom panel: spatially-augmented synthetic control (SA-SC). }
    \label{fig:plot-counties}
\end{figure}

All estimates were obtained using both the SC and SA-SC methods implemented under the SUTNVA assumption, which ensures that both approaches target the same causal estimand in the presence of local spatial interference. Figure \ref{fig:hw-definitions} presents estimates obtained under alternative heatwave definitions based on the 97th and 99th percentiles of the county-specific heat index distribution. These results illustrate the sensitivity of estimated heatwave effects to the choice of intensity threshold. Figure \ref{fig:five-years-maps} displays posterior relative risks aggregated within five-year periods. This temporal stratification highlights how the magnitude and spatial distribution of estimated effects have evolved over the study period. Figure \ref{fig:plot-counties} reports posterior mean relative risks and associated credible intervals for all Northeastern urban counties included in the analysis. This figure highlights county-specific heterogeneity and the degree of uncertainty in individual estimates. Figure \ref{fig:plot-counties-rollavg} presents analogous county-level posterior mean relative risks and associated credible intervals using a five-day rolling average applied to all days rather than imputing only zero-rate days. This figure illustrates the sensitivity of results to the treatment of zero rates, and we observe smaller estimated effects under this specification.

%\clearpage
%\subsection{Maps by five-year periods}

%\clearpage
%\subsection{County-level posterior estimates}

%\clearpage
%\subsection{Year-level posterior estimates}

%\begin{figure}[b]
%    \centering
%\includegraphics[width=\linewidth]{Figures/forest_plot_years.png}
 %   \caption{Posterior mean relative risk of heat-related hospitalizations during heatwaves versus non-heatwave days for all urban Northeastern U.S. county, depicted 2000-2019. Heatwaves are defined as periods of at least two consecutive days exceeding the 95th percentile of  county-specific annual heat index values. Estimates correspond to first-season heatwaves with donor pools restricted to non-exposed counties outside  treated neighborhoods. Top panel: standard synthetic control (SC). Bottom panel: spatially-augmented synthetic control (SA-SC).}
%    \label{fig:placeholder}
%\end{figure}

% \begin{figure}[h]
%     \centering
%     \includegraphics[width=\linewidth]{Figures/forest_plot_counties_rollavg.png}
%     \caption{Posterior mean relative risk of heat-related hospitalizations during heatwaves versus non-heatwave days for each urban Northeastern U.S. county considered in our study, 2000-2019, using a five-day right-aligned rolling average of daily hospitalization rates. Heatwaves are defined as periods of at least two consecutive days exceeding the 95th percentile of  county-specific annual heat index values. Estimates correspond to first-season heatwaves with donor pools restricted to non-exposed counties outside  treated neighborhoods. Top panel: standard synthetic control (SC). Bottom panel: spatially-augmented synthetic control (SA-SC). }
%     \label{fig:plot-counties-rollavg}
% \end{figure}

%% file: Tables/notation_table.tex
\begin{table}[htbp]
\begin{center}
\setlength{\tabcolsep}{3.5pt}
\resizebox{\linewidth}{!}{
\begin{tabular}{clclcl}
\toprule
\textbf{Symbol} & \textbf{Description} & \textbf{Symbol} & \textbf{Description} & \textbf{Symbol} & \textbf{Description} \\
\midrule
\multicolumn{6}{l}{\textbf{\textit{Basic Indices and Dimensions}}} \\
$i$ & Unit index & $N$ & Number of units & $T_0$ & Heatwave onset time \\
$j$ & Neighbor index & $t$ & Time index & $\ell$ & Heatwave duration \\
$k$ & Number of nearest neighbors & $T$ & Total time periods & $p$ & Covariate dimension \\
$r$ & Percentile threshold & $s_{ij}$ & Degree of separation & $s_0$ & Boundary for SUTNVA \\
\midrule
\multicolumn{6}{l}{\textbf{\textit{Variables and Data}}} \\
$Y_{it}$ & Health outcome & $H_{it}$ & Temperature/heat index & $\mathbf{X}_{it}$ & Covariates vector \\
$Y_{it}(z)$ & Potential outcome & $q_i^r$ & $r$-th percentile temp. & $Z_{it}$ & Treatment indicator \\
$Y_{it}(z,m)$ & Outcome under SUTNVA & $\mathcal{Z}_t$ & Treated units at $t$ & $C_i$ & Community indicator \\
\midrule
\multicolumn{6}{l}{\textbf{\textit{Sets and Spaces}}} \\
$\mathcal{N}_0^t$ & Donor pool at $t$ & $\mathcal{T}^-$ & Pre-treatment times & $N_i^{(k)}$ & $k$-degree neighbors \\
$\mathcal{N}_0^{*t}$ & Spatial donor pool & $\mathcal{T}^+$ & Post-treatment times & $\Delta^{n}$ & Simplex \\
\midrule
\multicolumn{6}{l}{\textbf{\textit{Parameters - Regression Models}}} \\
$\alpha$ & Intercept/fixed effects & $\beta$ & Treatment effect & $\boldsymbol{\gamma}$ & Covariate coefficients \\
$\alpha_i$ & Unit-specific effect & $\beta_i$ & Unit-specific effect & $\zeta_{it}$ & Poisson mean \\
\midrule
\multicolumn{6}{l}{\textbf{\textit{Parameters - Factor Models}}} \\
$\mu_i$ & Unit-specific effect & $\delta_t$ & Time effects & $f_t$ & Time-varying factors \\
$\lambda_i$ & Unit loadings & $\varepsilon_{it}$ & Idiosyncratic error & $u_{it}$ & Spatial error \\
\midrule
\multicolumn{6}{l}{\textbf{\textit{Functions}}} \\
$v(\cdot)$ & Smooth function & $h(\cdot)$ & Temperature function & $m(\cdot)$ & Exposure mapping \\
$g(\cdot)$ & Time spline & & & & \\
\midrule
\multicolumn{6}{l}{\textbf{\textit{Weights and Spatial Components}}} \\
$w_j$ & SC weights & $\omega_j^i$ & Bayesian weights & $\mathbf{W}_{geo}$ & Spatial weight matrix \\
$D_{ij}$ & Geographic distance & $s_{ij}$ & Degree of separation & $\mathbf{W}_{adj}$ & Binary adjacency matrix  \\
\midrule
\multicolumn{6}{l}{\textbf{\textit{Bayesian Parameters}}} \\
$\eta_j^i$ & Latent coefficients & $\varsigma$ & Spatial decay & $\sigma$ & Standard deviation \\
$\tau$ & Effect & $\pi(\cdot)$ & Posterior distribution & $\boldsymbol{\theta}$ & Parameter vector \\
\midrule
\multicolumn{6}{l}{\textbf{\textit{Simulation Parameters}}} \\
$\varpi_\alpha$ & SAR parameter & $\rho_u$ & Spatial correlation & $\chi$ & Spillover decay \\
$\kappa$ & Effect multiplier & $\psi_{jt}^s$ & Spillover effect & $\nu_\alpha$ & SAR innovations \\
\bottomrule
\end{tabular}}
\caption{\normalsize Notation summary.}
\label{tab:notation}
\end{center}
\end{table}

%% file: Tables/sim_sutva.tex
\begin{table}
    \centering
    \resizebox{\linewidth}{!}{
\begin{tabular}{llrrrrr}
\toprule
  & Method & Abs avg. Bias & RMSE pre & RMSE post & Coverage prob. & Avg. CI length\\
\midrule
No Spatial depend. - No spillover & SC & 0.820 & 1.311 & 1.455 & 0.936 & 3.744\\ \rowcolor{lightgray}
 & SA-SC & 0.718 & 1.194 & 1.354 & 0.948 & 3.484\\
No Spatial depend. - Spillover & SC & 0.829 & 1.313 & 1.443 & 0.937 & 3.753\\ \rowcolor{lightgray}
 & SA-SC & 0.894 & 1.172 & 1.806 & 0.899 & 3.636\\
Spatial depend. - No Spillover & SC & 0.830 & 1.334 & 1.474 & 0.931 & 3.770\\ \rowcolor{lightgray}
 & SA-SC & 0.694 & 1.126 & 1.322 & 0.942 & 3.389\\
Spatial depend. - Spillover & SC & 0.833 & 1.295 & 1.490 & 0.933 & 3.769\\ \rowcolor{lightgray}
 & SA-SC & 0.840 & 1.094 & 1.683 & 0.902 & 3.530\\
\bottomrule
\end{tabular}
    }
    \vspace{0.1cm}
    \caption{Monte Carlo performance under SUTVA assumption across 100 replications for causal inference methods.  SC refers to \textit{synthetic control} while SA-SC refers to \textit{spatially augmented synthetic control}.}
    \label{tab:sim_res_sutva}
\end{table}

%% file: Tables/descriptive_table.tex
\begin{table}[h]
    \centering
    
\begin{tabular}{lrr}
\toprule
Variable & Urban & Non-urban \\
\midrule

\textbf{Population characteristics} & & \\
\rowcolor{lightgray}\hspace{0.25cm}Beneficaries & 8,022,444 & 6,086,858 \\
\hspace{0.25cm}Beneficiary-years & 60,994,233 & 47,856,005 \\
\rowcolor{lightgray}\hspace{0.25cm}Daily hospitalization rate (per 100,000) & 9.89 & 8.29 \\
\hspace{0.25cm}Medicare population (thousands) & 13.8 & 70.9 \\
\rowcolor{lightgray} \hspace{0.25cm}County population (thousands) & 795.4 & 124.3 \\
\hspace{0.25cm}Number of counties & 43 & 174 \\

\midrule
\textbf{Age (\%)} & & \\
\rowcolor{lightgray}\hspace{0.25cm}65--74 years & 49.9 & 50.1 \\
\hspace{0.25cm}75--84 years & 32.6 & 33.2 \\
\rowcolor{lightgray}\hspace{0.25cm}85--94 years & 15.3 & 15.0 \\
\hspace{0.25cm}$\geq$95 years & 2.2 & 1.8 \\

\midrule
\textbf{Sex (\%)} & & \\
\rowcolor{lightgray}\hspace{0.25cm}Female & 58.5 & 56.8 \\

\midrule
\textbf{Race (\%)} & & \\
\rowcolor{lightgray}\hspace{0.25cm}White & 83.4 & 96.0 \\
\hspace{0.25cm}Black & 8.2 & 1.3 \\

\midrule
\textbf{Socioeconomic indicators} & & \\
\rowcolor{lightgray}\hspace{0.25cm}Higher education (\%) & 10.5 & 7.25 \\
\hspace{0.25cm}Poverty (\%) & 11.1 & 11.8 \\
\rowcolor{lightgray}\hspace{0.25cm}Owner occupied housing (\%) & 58.6 & 57.7 \\
\hspace{0.25cm}Home value (thousand USD) & 342.9 & 177.0 \\
\rowcolor{lightgray}\hspace{0.25cm}Household income (thousand USD) & 71.6 & 54.6 \\

\bottomrule
\end{tabular}

\vspace{0.1cm}
\caption{Descriptive statistics for urban and non-urban counties. Values represent averages across counties and years from 2000 to 2019. Counties are classified as urban or non-urban based on metropolitan area status and population greater than 200,000, and beneficiaries contribute to each group according to the county in which they reside in a given year. Beneficiaries can enter or exit Medicare during the study period, and some individuals move between urban and non-urban counties, so the two groups do not represent fixed cohorts. Population characteristics summarize the size of the Medicare population, the number of beneficiary-years, and the average daily hospitalization rate. Individual-level demographic covariates include age, sex, and reported race. Socioeconomic indicators were linked from the American Community Survey at the county level. Higher education attainment reflects the percentage of the population with a higher education degree, the poverty rate captures the percentage of the population living below the poverty line, owner-occupied housing represents the percentage of housing units occupied by their owner, median home value refers to the median value of owner-occupied properties, and household income summarizes median income levels within each county.}
\label{tab:urban_rural_desc}
\end{table}